\documentclass[preprint]{aastex}

\shorttitle{KIC 5738698}
\shortauthors{Matson et al.}

\usepackage{natbib}
\usepackage{graphicx}
\usepackage{url}
\usepackage{relsize}

\newcommand{\noprint}[1]{}
\newcommand{\figsetstart}{{\bf Fig. Set} }

\newcommand{\figsetgrpstart}{}
\newcommand{\figsetgrpend}{}
\newcommand{\figsetnum}[1]{{\bf #1.}}
\newcommand{\figsettitle}[1]{ {\bf #1} }
\newcommand{\figsetgrpnum}[1]{\noprint{#1}}
\newcommand{\figsetgrptitle}[1]{\noprint{#1}}
\newcommand{\figsetplot}[1]{\noprint{#1}}
\newcommand{\figsetgrpnote}[1]{\noprint{#1}}

\begin{document}

\title{Fundamental Parameters of Kepler Eclipsing \\
	Binaries. I. KIC 5738698}

\author{Rachel A. Matson\altaffilmark{1}, Douglas R. Gies, Zhao Guo}
\affil{{\it Center for High Angular Resolution Astronomy and Department of Physics and Astronomy, Georgia State University, P.O. Box 5060, Atlanta, GA 30302-5060, USA}}
\email{rmatson@chara.gsu.edu, gies@chara.gsu.edu, guo@chara.gsu.edu}

\author{Jerome A. Orosz}
\affil{{\it Department of Astronomy, San Diego State University, San Diego, CA 92182-1221, USA}}
\email{jorosz@mail.sdsu.edu}

\altaffiltext{1}{Visiting astronomer, Kitt Peak National Observatory, National Optical Astronomy Observatory, which is operated by the Association of Universities for Research in Astronomy (AURA) under a cooperative agreement with the National Science Foundation.}

\begin{abstract}

Eclipsing binaries serve as a valuable source of stellar masses and radii that inform stellar evolutionary models and provide insight into additional astrophysical processes. The exquisite light curves generated by space-based missions such as {\it{Kepler}} offer the most stringent tests to date. We use the {\it{Kepler}} light curve of the 4.8~day eclipsing binary KIC~5739896 with ground based optical spectra to derive fundamental parameters for the system. We reconstruct the component spectra to determine the individual atmospheric parameters, and model the {\it{Kepler}} photometry with the binary synthesis code ELC to obtain accurate masses and radii. The two components of KIC~5738698 are F-type stars with $M_{1} = 1.39\pm0.04M_{\odot}$, $M_{2} = 1.34\pm0.06M_{\odot}$, and $R_{1} = 1.84\pm0.03R_{\odot}$, $R_{2} = 1.72\pm0.03R_{\odot}$. We also report a small eccentricity ($e \lesssim 0.0017$) 
and unusual albedo values that are required to match the detailed shape of the {\it{Kepler}} light curve. Comparison with evolutionary models indicate an approximate age of 2.3~Gyr for the system. 

\end{abstract}


\section{Introduction}
\label{sec:intro}

NASA's {\it{Kepler}} Mission to search for transiting exoplanets provided high-precision nearly uninterrupted photometry of over 160,000 objects that helped revolutionize not only planetary, but stellar astrophysics as well. The diverse causes of photometric variability in stars, including multiplicity and pulsations, can now be studied with unprecedented detail. More than 2600 eclipsing binaries, a vital source of stellar mass and radius measurements, have been discovered with {\it{Kepler}} (\citet{Prsa:aj2011a}; \citet{Slawson:aj2011a}; \citet{Kirk:ArXiv2015a}). Individual systems consisting of low-mass stars \citep{Ofir:mnras2012a, Bass:apj2012a}, red-giants \citep{Gaulme:apj2013a, Beck:aap2014a}, $\delta$ Scuti/$\gamma$ Doradus pulsators \citep{Southworth:mnras2011a, Lehmann:aap2013a, Debosscher:aap2013a, Maceroni:aap2014a}, and heartbeat stars \citep{Hambleton:mnras2013a} have provided accurate mass and radius estimates in addition to probing our understanding of proximity effects and intrinsic stellar variability. 

The sensitivity and long time span of {\it{Kepler}} observations have also opened up the search for triple stellar systems via multiple transits and periodic variations in the eclipse times. Theoretical studies suggest that a significant number of close binaries have a distant tertiary companion which plays an important role in the evolution of the inner binary through the exchange of angular momentum \citep{Eggleton:apss2006a,Fabrycky:apj2007a,Naoz:apj2014a}. Observational evidence of such systems has been found by \citet{Pribulla:aj2006a}, \citet{Tokovinin:aap2006a}, \citet{Raghavan:apjs2010a}, and \citet{Zakirov:KPCB2010a}. {\it{Kepler}} has managed to increase the discovery of close binaries with tertiary companions \citep{Carter:Science2011a, Lee:apj2013a, Rappaport:apj2013a, Borkovits:mnras2013a, Borkovits:mnras2015a, Lee:aj2015a} and provide insight into the stability, dynamics, and evolution of such systems.

With the launch of {\it{Kepler}}, we began a Guest Observer program to measure eclipse timing variations (ETV) for a sample of `pristine' {\it{Kepler}} light curves to search for tertiary companions to close binaries \citep{Gies:aj2012a}. These systems are detached or semi-detached with deep eclipses ($>$\,$0.2$ mag), primary effective temperatures between 5200-11000\,K, and orbital periods less than six days. Eclipse timing results through quarter 9 found long term variations in 14 of the binary systems (34\%) that may be due to tertiary companions. An update to this work, including all 18 quarters of data, identified seven probable triple systems and seven additional systems that may be triples with orbits longer then the {\it{Kepler}} baseline \citep{Gies:aj2015a}. In order to constrain the masses of tertiary companions and to characterize fully the binaries in these 41 systems, we have also collected ground based optical spectra and derived radial velocities and spectroscopic orbits (R.\,Matson et al.,~in prep.).

Here we present our methods and the results of a full analysis of the light curve and spectroscopy of KIC~5738698, a detached eclipsing binary consisting of two nearly identical F-type stars orbiting with a period of 4.8\,days. KIC~5738698 was detected in the HATNET \citep[199-19185;][]{Hartman:aj2004a} and ASAS \citep[J195853+4054.2;][]{Pigulski:actaa2009a} surveys. It is listed in the Kepler Input Catalog (KIC)/Kepler Target Catalog (KTC) as having $T_{\rm{eff}}  = 6210$\,K, $\log g = 4.259$, $\log \rm{[Fe/H]} = -0.490$, and $R = 1.317R_\odot$, while \citet{Armstrong:mnras2014a} derived $T_{\rm{eff,1}} = 6578\pm358$\,K, $T_{\rm{eff,2}} = 6519\pm555$\,K, $R_2/R_1 = 0.83\pm0.32$ and $T_{\rm{1}}/T_{\rm{2}} = 0.9905$ in their catalog of temperatures for {\it{Kepler}} eclipsing binary stars. 

While the eclipse timing measurements of KIC~5738698 \citep{Gies:aj2015a} do not show any evidence of a third star, we chose this system to highlight our methods as well as the additional considerations that arise when modeling {\it{Kepler}} photometry. We also use the exquisite {\it{Kepler}} data and our spectroscopy to measure accurately the masses and radii of KIC~5738698, contributing to our knowledge of detached systems with accurately known fundamental parameters that serve as valuable tests of stellar evolution \citep{Torres:aapr2010a}, specifically in the mass/temperature regime where convective cores begin to develop and affect the observational properties of the stars \citep{Clausen:aap2008a}.

We discuss our observations in Section~\ref{sec:obs}, followed by the determination of radial velocities and atmospheric parameters from reconstructed spectra in Section~\ref{sec:spec}. In Section~\ref{sec:elc} the binary modeling method is described, including details of a circular baseline model, efforts to minimize the residuals, and parameter uncertainties. Section~\ref{sec:evol} compares our results to theoretical predictions of several evolutionary models, followed by a short summary in Section~\ref{sec:sum}.

\section{Observations}
\label{sec:obs}
\subsection{{\it{\textbf{Kepler}}} Photometry and Orbital Ephemeris}
\label{subsec:ephem}

Data from the {\it{Kepler}} satellite are obtained by the on-board photometer using a set integration time of $6.54$\,s (including readout), which is co-added into long and short cadence data sets of $29.4244$\,m and $58.8488$\,s, respectively \citep{Gilliland:apjl2010a}. Data are combined in quarters corresponding to spacecraft rolls approximately every three months \citep[to maintain optimal solar panel illumination;][]{Haas:apjl2010a}. For KIC~5738698 there are 18 quarters of long cadence (Q0-17; 2009 May 2 -- 2013 May 8) and one month (Q4.1; 2009 Dec 19 -- 2010 January 19) of short cadence data. 

\begin{figure}
\begin{center}
\includegraphics[scale=0.6,angle=90]{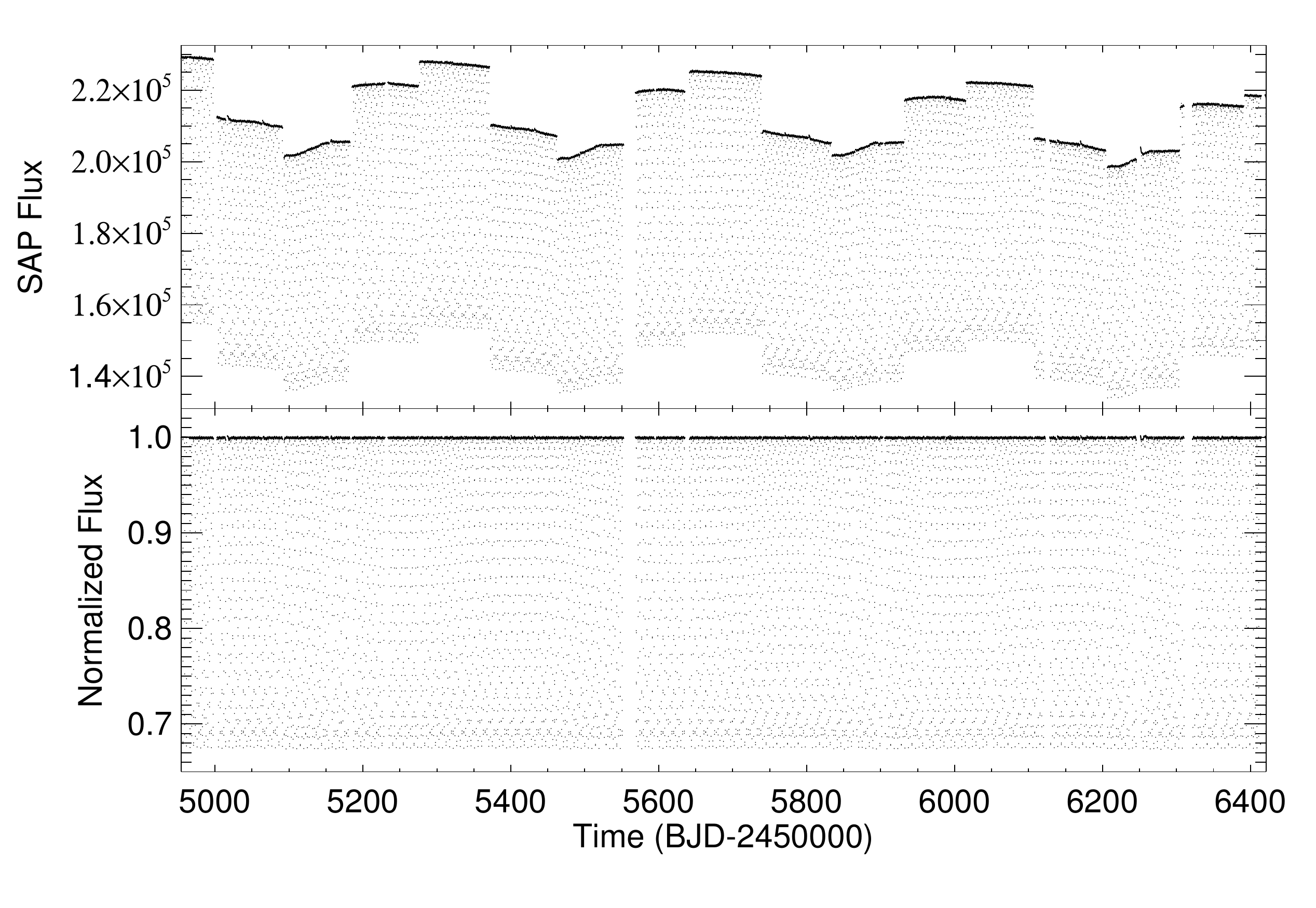}
\caption{Top: the SAP light curve of KIC~5738698 from Q0 to Q17, showing the changes in flux levels between quarters. Bottom: the detrended and normalized light curve.}
\label{fig:kepdata}
\end{center}
\end{figure}

We use the Simple Aperture Photomtery (SAP) light curves output by the {\it{Kepler}} data processing pipeline, correcting for varying flux levels within quarters by binning the data to give a minimum scatter in out-of-eclipse phases and fitting a cubic spline through the mean of the upper 50\% of each section. Each quarter was then divided by the spline fit before being combined into a single light curve \citep[as in][]{Gies:aj2012a}. Outliers greater than 3$\sigma$ from the out-of-eclipse median magnitude were removed. The SAP and detrended light curves are shown in Figure~\ref{fig:kepdata}. 

While {\it{Kepler}} data is affected by a variety of systematic trends and instrumental artifacts \citep{Kinemuchi:pasp2012a}, fitting a cubic spline through binned segments adequately removes the large scale trends found within each quarter. This method does not completely account for jumps, drifts, and outliers, but does mitigate their effects while maintaining the binary signature. Any remaining artifacts should be randomized and merely add to the overall scatter of the residuals as we phase fold the light curve for all subsequent analysis.

An updated ephemeris for KIC~5738698 was determined from eclipse templates made with the binned and folded {\it{Kepler}} light curve in Gies et al.~(2015). A period of $4.80877396\pm 0.000000035$ days was adopted from the average of the individual periods of the primary and secondary eclipses. The epoch of mid-eclipse of the primary is $2455692.3348702\pm0.000002$ (BJD). More details and a list of improved periods and epochs for all 41 systems can be found in \citet[][2015]{Gies:aj2012a}.

\subsection{Ground-based Spectroscopy}
\label{subsec:specobs}

Thirteen moderate resolution spectra of KIC~5738698 were obtained between 2010 June and 2011 September at Kitt Peak National Observatory (KPNO) and the Anderson Mesa Station of Lowell Observatory. Observations at KPNO were made with the 4\,m Mayall telescope and R-C Spectrograph using the BL380 grating (1200 grooves mm$^{-1}$) in second order, providing wavelength coverage of $3930-4610$\AA~with an average resolving power of $R=\lambda/\delta\lambda\approx6200$. Calibration exposures using HeNeAr lamps were taken either immediately before or after each exposure, and bias and flat-field spectra were obtained nightly.

Observations of KIC~5738698 at Lowell Observatory were conducted on the 1.8\,m Perkins telescope with the DeVeny Spectrograph. Using a 2160 grooves mm$^{-1}$ grating, we obtained a resolving power of $R=\lambda/\delta\lambda\approx6000$ over the wavelength range $4000-4530$\AA. Calibration exposures with HgNeArCd Pen-Ray lamps were taken before or after each exposure, and bias and flat-field spectra were taken nightly.

All spectra were reduced, extracted, and wavelength calibrated using the corresponding comparison lamp spectra and standard IRAF\footnote{IRAF is distributed by the National Optical Astronomy Observatory, which is operated by the Association of Universities for Research in Astronomy (AURA), Inc., under cooperative agreement with the National Science Foundation.} procedures. However, the comparison lamps for the Lowell DeVeny spectrograph only produce three measurable emission lines in the wavelength region $4000-4530$\AA~(Hg I $\lambda4046, 4077, 4358$). Therefore, standard velocity stars were observed $2-4$ times per night to aid in the determination of the dispersion solution. Model spectra from the UVBLUE\footnote{\url{http://www.inaoep.mx/~modelos/uvblue/download.html}} libraries of \citet{Rodriguez-Merino:apj2005a} interpolated to the appropriate temperatures and gravities were transformed to the topocentric velocity of each standard star and convolved for instrumental broadening. We used the comparison lamp exposures taken with each standard velocity star to determine an initial fit of wavelength to pixel number. The observed spectra and appropriate model were then cross-correlated in 40 sub-regions across the spectrum to get the mean pixel and wavelength values of each region. These values were then fit with a cubic polynomial to remove any systematic effects across the chip. Finally, dispersion corrections were applied to the science spectra observed nearest in time to each standard, based on the cubic polynomial 
and individual pixel shifts determined from the comparison lamp spectra. For KIC~5738698 we include one spectrum taken at Lowell Observatory in 2010 July for which we used the velocity standard HD~187691 ($T_{\rm{eff}} = 6107$\,K, $\log g = 4.30$, \citealt{Cenarro:mnras2007a}; $V_r = -0.15\pm0.40$ km s$^{-1}$, \citealt{Molenda-Zakowicz:actaa2007a}). 

After wavelength calibration, all spectra were rectified to a unit continuum and transformed to a common heliocentric wavelength grid in log $\lambda$ increments.

\section{Spectral Analysis}
\label{sec:spec}
\subsection{Radial Velocities}
\label{subsec:rv}


Radial velocities were measured with a two-dimensional cross-correlation technique, based on the method used in PROCOR by R.W. Lyons \citep[see][]{Gies:apjs1986a}, employing template spectra to determine the velocity separation of the secondary component relative to the primary and the absolute velocity of the primary. Separate templates for the primary and secondary were selected from the UVBLUE grid of high resolution model spectra, 
which is based on LTE calculations using the ATLAS9 and SYNTHE codes of R.\,L.~Kurucz \citep{Rodriguez-Merino:apj2005a}. The synthetic spectra were chosen based on initial estimates of temperatures, gravities, projected rotational velocities, and relative flux contributions of each star calculated from the {\it{Kepler}} Eclipsing Binary Catalog results of \citet{Slawson:aj2011a}. 
Each spectrum was rebinned onto the observed wavelength grid and convolved for the projected rotational velocity ($v \sin i$) and instrumental broadening. Solar metallicity was assumed throughout. 

We estimated preliminary orbital elements from the parameters of \citet{Slawson:aj2011a}, and then used the times of observation to predict radial velocities that determined trial velocity separations for the primary and secondary components.
These trial separations were used to make a series of composite model spectra over a grid of separations that were cross-correlated with each observed spectrum. We plotted the maximum of each cross-correlation function against the corresponding trial separation and determined the optimal separation by fitting a parabola through the points and selecting the separation at the interpolated global maximum.
A final cross-correlation was performed using this separation to get the absolute velocity of the primary which, when combined with the separation, gave us the secondary velocity. After deriving atmospheric parameters from the tomographically reconstructed spectra (\S\,\ref{subsec:tomog}), the templates were updated and radial velocities re-derived. The final radial velocities for KIC~5738698 are listed in Table~\ref{tab:rvs}, along with the date of observation in Heliocentric Julian days, orbital phase, uncertainty $\sigma$, and observed minus calculated ($O-C$) residuals from the spectroscopic fit (\S\,\ref{subsec:sporb}). Orbital phase is determined relative to T$_0$, taken to be the epoch of primary eclipse for this paper.


\subsection{Orbital Solution}
\label{subsec:sporb}

We determined orbital elements for KIC~5738698 using a nonlinear, least-squares fitting routine \citep{Morbey:pasp1974a}. The period was held fixed to the value obtained from the eclipse timings (\S\,\ref{subsec:ephem}), while the epoch was allowed to vary. The radial velocities were weighted by the inverse square of the uncertainties, with the exception of an anomalous measurement of the primary component from Lowell Observatory which was zero weighted and therefore omitted from the fitting process. While we would expect the radial velocity measurement of the secondary component to be similarly affected, it appears comparable to the other measurements near the same phase and we therefore chose to include it after verifying that doing so did not alter the orbital solution. Although initial fits indicated that the orbit of KIC~5738698 is circular, as expected for short period systems, we fit the radial velocities with both circular and eccentric orbits. The statistical significance of each fit was evaluated according to \citet{Lucy:aj1971a} and \citet{Lucy:aap2013a}, but the eccentric orbit failed to improve the fit in both cases and we therefore adopt the circular orbit for now (see \S \ref{subsubsec:ecc}).

\begin{figure}
\begin{center}
\includegraphics[scale=0.6,angle=90]{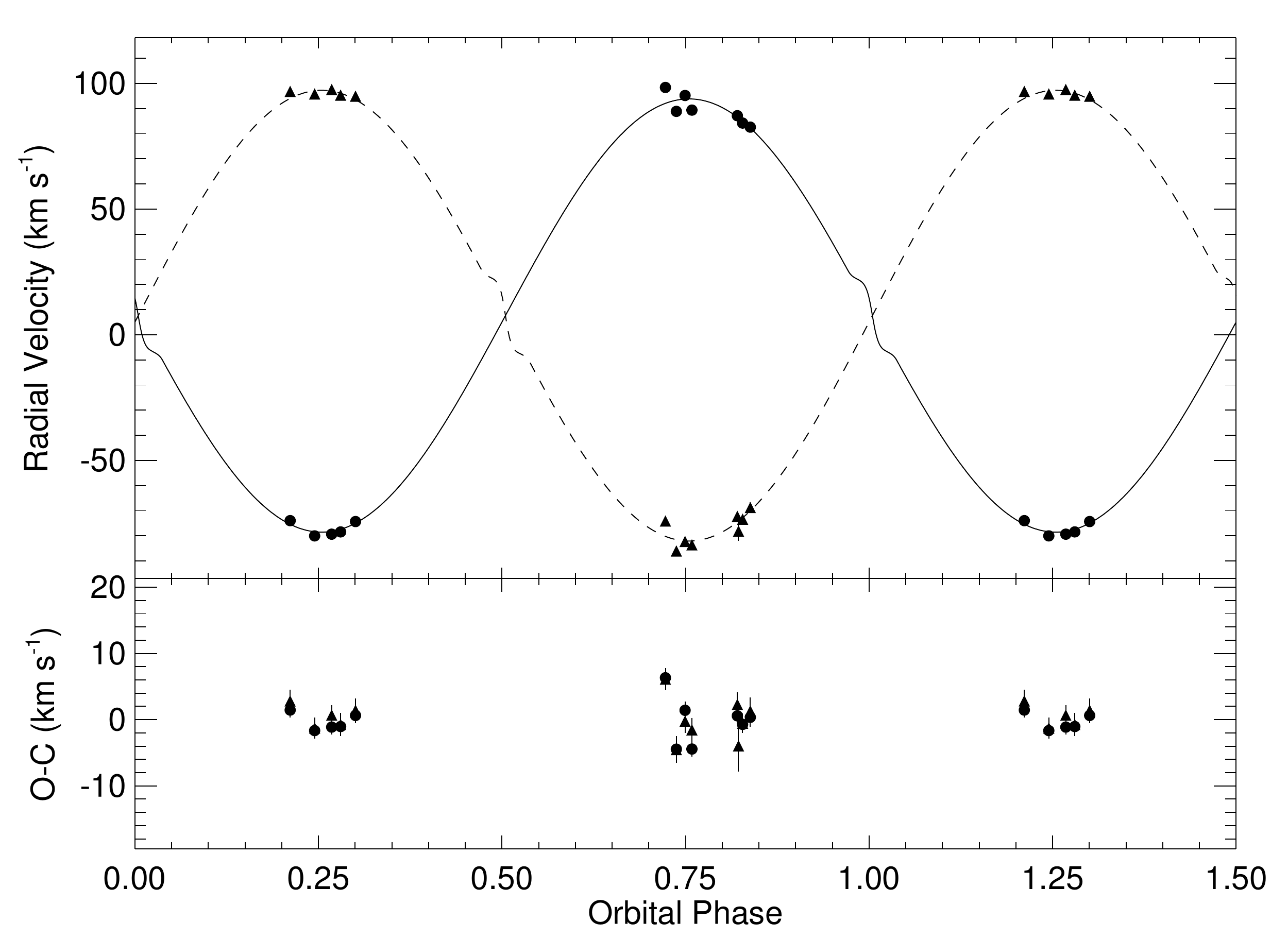}
\caption{Top: radial velocity curves of the primary (filled circles) and secondary (filled triangles) of KIC~5738698 and the best fitting ELC model. Phase zero corresponds to the time of primary eclipse. Bottom: residuals for the fits to the primary and secondary velocities.}
\label{fig:rv}
\end{center}
\end{figure}

Parameters for the circular fit to the orbit of KIC~5738698, including the period ($P$), time of primary eclipse ($T_0$), systemic velocity ($\gamma$), and velocity semi-amplitude of the primary ($K_1$) and secondary ($K_2$) are given in Table~\ref{tab:sporb} as well as the derived mass ratio ($q=M_2/M_1$) and $a\sin i$. The primary and secondary radial velocities were fit separately, which allows us to check the consistency of the fits as well as our derived radial velocities. The epochs determined from both components agree within uncertainties ($T_{0_1} = 55692.330\pm0.032$, $T_{0_2} + \frac{P}{2} = 55692.340\pm0.028$) and are consistent with the epoch from the eclipse timings, though less precise. Similarly, the systemic velocities from each fit ($\gamma_1 = 7.4\pm0.9$~km s$^{-1}$, $\gamma_2 = 8.1\pm0.8$~km s$^{-1}$) are consistent within the uncertainties and we adopt the weighted mean (see Table~\ref{tab:sporb}). The radial velocities, residuals, and an updated orbital solution (as described in \S\,\ref{subsec:elcrv}) are plotted in Figure~\ref{fig:rv}.

\subsection{Spectral Reconstruction and Atmospheric Parameters}
\label{subsec:tomog}

\begin{figure}
\begin{center}
\includegraphics[scale=0.7,angle=90]{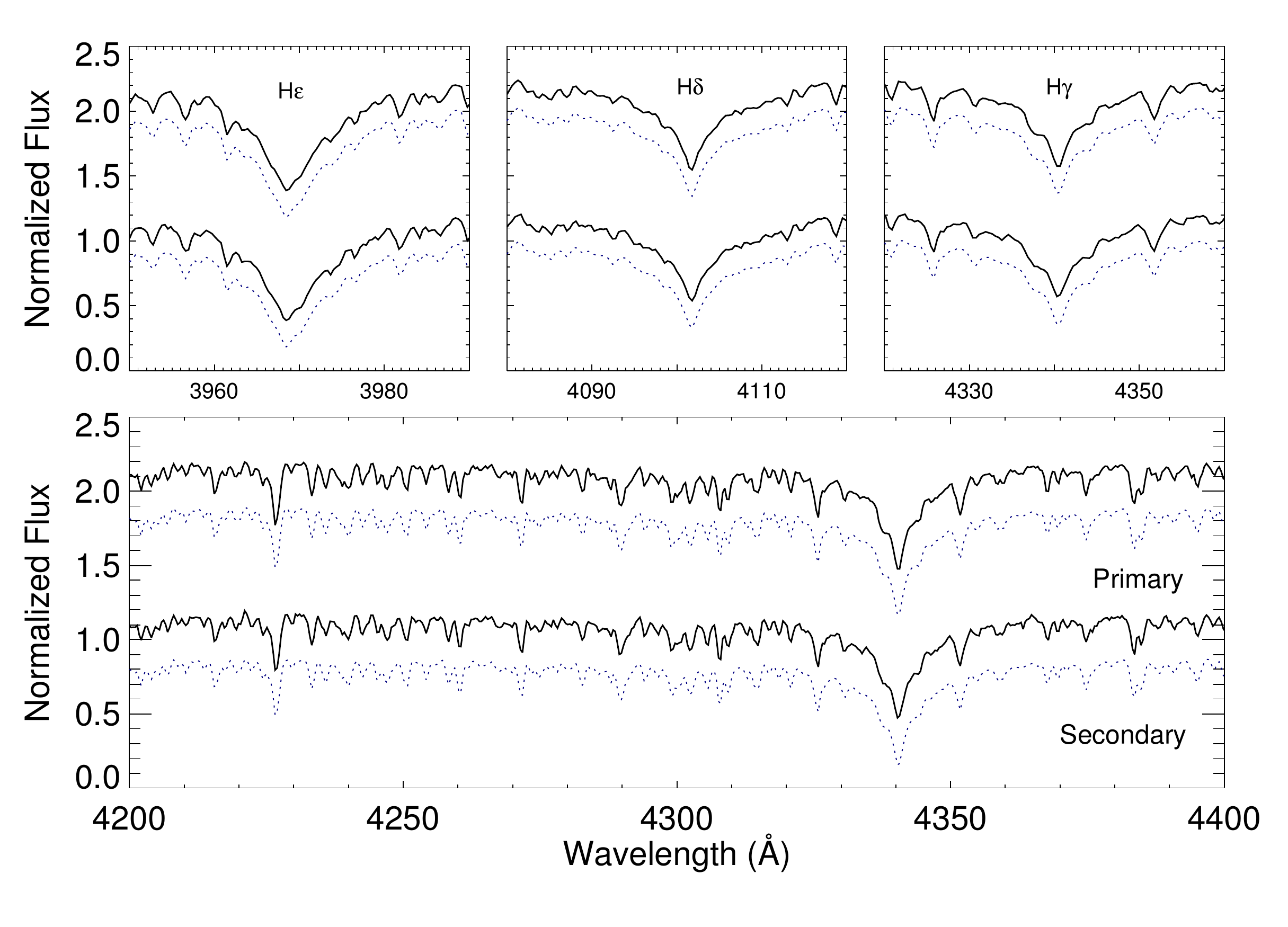}
\caption{A portion of the reconstructed spectra of the primary (upper) and secondary (lower) components of KIC~5738698 based on 13 moderate resolution optical spectra. Corresponding model spectra from the UVBLUE grid are shown as (blue) dashed lines (offset by -0.3 normalized flux units). The upper (3) panels depict the reconstructions of the individual hydrogen Balmer lines, which are particularly sensitive to temperature in this spectral region and (H$\delta$ and H$\gamma$) were used to constrain the effective temperatures of each component.}
\label{fig:tomog}
\end{center}
\end{figure}

We used the Doppler tomography algorithm of \citet{Bagnuolo:apj1994a} to reconstruct the primary and secondary spectra of KIC~5738698. Using the composite spectra, radial velocities, and flux ratio of the primary and secondary, this method iteratively shifts and adds flux from each component in proportion while making small corrections via a least-squares technique. Reconstructed spectra of the primary and secondary are shown in Figure~\ref{fig:tomog} along with model spectra as described below.

Using the individual reconstructed spectra we can derive atmospheric parameters for each star by comparing them with model spectra. We begin by comparing the reconstructed spectra with UVBLUE models based on initial parameter estimates from the {\it{Kepler}} Eclipsing Binary Catalog (as in our radial velocity determination) over a range of rotational broadenings to determine $v$ sin $i$ for the primary and secondary. By minimizing a chi-squared goodness-of-fit statistic for the metallic lines, we find projected rotational velocities of $18\pm16$ km s$^{-1}$ for the primary and $21\pm10$ km s$^{-1}$ for the secondary of KIC~5738698, consistent with the estimated synchronous rate of 18.6 km s$^{-1}$. While these errors indicate large uncertainties in our derived rotational velocities, they are expected as the $v \sin i$ values are at the limits of our spectral resolution. Based on our resolving power of ${\lambda}/{\Delta\lambda}\sim6200$, we can only reliably measure $v \sin i$ as small as $c/2R\sim 24$ km s$^{-1} $. However, as the $v \sin i$ values are near the expected synchronous rotation rate we adopt them and fix $\log g$ (initially to the values from the second release of the {\it{Kepler}} Eclipsing Binary Catalog, then according to our light curve solution \S\,\ref{subsec:elclc}) for the model comparisons. Fixing $\log g$ in this way helps mitigate the known degeneracies between $\log g$, $T_{\mathrm{eff}}$, and $\log Z$ \citep{Torres:apj2012a}.

Because the properties of the tomographic reconstruction depend upon the assumed flux ratio $r$ and (slightly) on the metallicity, to determine the best solution and remaining stellar parameters we make separate fits to the primary and secondary over a grid of $ r = F_{2}/F_{1}$ and $\log Z/Z_\odot$ values. 
Tomography is repeated at each grid point and best-fit temperatures are determined via a least-squares fitting routine (using lines most sensitive to temperature, specifically H$\delta$ and H$\gamma$ for KIC~5738698). The best $\log Z/Z_\odot$ and $r = F_{2}/F_{1}$ fits for the primary and secondary individually form two `valleys' of minimum chi-squared as a function of $\log Z/Z_\odot$ and $r$, and we select the intersection of these two valleys so that the metallicity and flux ratio are the same for both stars. The final temperatures are then derived for each component using the $\log Z/Z_\odot$ and $r$ at this consistent minimum. Table~\ref{tab:atm} gives the stellar parameters derived from the best-fit models to the tomographic reconstructions for the primary and secondary. 

As seen in Figure~\ref{fig:tomog}, the line depths and overall appearance of the model spectra are in very good agreement with the reconstructed spectra. The Balmer lines are the most temperature sensitive features in our wavelength range and are extremely well fit in both depth and width. Other lines sensitive to temperature and metallicity such as Ca I $\lambda$4226, Fe I $\lambda\lambda$4046, 4271, and 4383 \citep{Gray:2009} also show consistent fits.  

Parameter uncertainties were computed via bootstrapping, in which we randomly resampled the input spectra for the tomography grids and used the standard deviation of 500 resampled solutions as the uncertainty. In addition, we also created a simulated stack of model spectra with the same Doppler shift sampling as the observations based on the derived stellar parameters and characteristic signal-to-noise levels. We then performed tomography and the grid search on the simulated spectra to determine how much the derived parameters differ from those used to create the model. Quoted parameter uncertainties (see Table~\ref{tab:atm}) were adopted from the technique that gave the larger estimated uncertainty or one-tenth the UVBLUE grid step size  of $\Delta T_{\rm{eff}} = 500$\,K and $\Delta \log Z = 0.5$\,dex.

\section{Binary Star Modeling}
\label{sec:elc}


The light curve and radial velocities of KIC~5738698 were modeled using the Eclipsing Light Curve (ELC) code of \citet{Orosz:aap2000a}. The code fits for a variety of binary star parameters using optimizers, such as the genetic algorithm employed in this work, to determine a best-fit model based on an overall chi-squared goodness-of-fit. ELC uses Roche geometry and specific intensities computed from PHOENIX model atmospheres \citep{Hauschildt:apj1997a} to determine the flux by numerically integrating over surface tiles.

\subsection{Radial Velocity Modeling}
\label{subsec:elcrv}

ELC is designed to simultaneously model radial velocities and photometric data in multiple bandpasses; however when fitting data with the precision of the {\it{Kepler}} data, the combined solutions converge where the radial velocities are not well fit due to the mismatch in the relative weights of the spectroscopic and photometric data \citep{Bass:apj2012a}. Because of this issue with the relative weights and the fact that KIC 5738698 is a well detached (nearly) circular system, we chose to fit the light curve and radial velocities separately (as done by \citealt{Bass:apj2012a}, \citealt{Sandquist:aj2013a}, \citealt{Jeffries:aj2013a}). While information on $e$ and $\omega$ is contained in both the radial velocity and light curves and they should usually be modeled simultaneously, the very small eccentricity we detect in the light curve (\S\,\ref{subsubsec:ecc}) cannot be constrained by the radial velocity curves to the precision at which we detect it in the light curve, so a constrained fit was made.

Although the relative weights of the radial velocities and {\it{Kepler}} data prevent a simultaneous solution, we did fit the radial velocities with ELC using the results of our previous spectroscopic orbit, tomographic reconstruction, and initial parameter estimates from the {\it{Kepler}} Eclipsing Binary Catalog. This second spectroscopic fit not only verifies the consistency of the two methods but also employs the more complete physical descriptions of the two stars in ELC, which can account for offsets between the center of light and center of mass of each component as in the Rossiter-McLaughlin and reflection effects. For the ELC model we allowed the 
 mass ratio ($q$), velocity semi-amplitude of the primary ($K_1$), and velocity zero point ($\gamma$) to vary. The period was again fixed to the value derived from eclipse timings in \citet{Gies:aj2015a} and a circular orbit was assumed initially. A revised (eccentric) radial velocity model was determined once the light curve parameters were derived (\S\,\ref{subsec:resid}) and can be seen in Figure~\ref{fig:rv}. The associated orbital parameters are given in Table~\ref{tab:sporb}, which shows that the results of this model agree quite well with our previous spectroscopic orbital solution. 
A slight discrepancy, though still within the uncertainties, occurs in the systemic or zero point velocity, which is not surprising as the spectroscopic solution determined $\gamma$ separately for each star, and we report the weighted mean of the two values (\S\,\ref{subsec:sporb}). One sigma uncertainties were determined by collapsing the $n$-dimensional $\chi^2$ function from ELC onto each parameter and determining where the lower envelope is equal to $\chi_{min}^2+1$ (see \S\,\ref{subsec:uncert}).

\subsection{Light Curve Modeling}
\label{subsec:elclc}

\subsubsection{Circular Orbit Model Parameters}
\label{subsubsec:lcmod}

Light curve models of KIC~5738698 were computed with ELC using all 18 quarters of {\it{Kepler}} long cadence data. We began by fixing the orbital period to $4.80877396$\,d from our eclipse timing results \citep{Gies:aj2015a} and assuming a circular orbit. The value of T$_0$ from the eclipse timings was initially used as the time of primary eclipse ($T_{\rm{conj}}$ in ELC), but better convergence was reached when it was allowed to vary slightly. The primary effective temperature was fixed according to the tomography results (\S\,\ref{subsec:tomog}), while the parameters from the ELC radial velocity fit ($q, K_1$, and $\gamma$) were held constant as they have little influence on the light curve solution.

Our baseline model had five free parameters: inclination ($i$), temperature ratio ($T_{2}/T_{1}$), fractional radii ($R_{1}/a$ and $R_{2}/a$), and time of primary eclipse ($T_{\rm{0}}$). 
We set ELC to use the included model atmosphere table to describe the variation of local intensities with emergent angle in the {\it{Kepler}} bandpass. This negates the need for limb darkening coefficients, except in the computation of the reflection effect, for which we used a logarithmic law \citep[which provides a better match in the optical for stars cooler than 9000\,K;][]{Prsa:apj2005a} and coefficients from \citet{Howarth:mnras2011a}. 
The bolometric albedo was set to 0.5 for convective envelopes \citep{Rucinski:actaa1969a} and the gravity darkening exponents were set internally, based on the input effective temperatures, according to \citet{Claret:aap2000a}. 
The model intensity is integrated over a grid of equal angle increments corresponding to 60 points in latitude and 80 points in longitude on the surface of the star.

\subsubsection{Features of {\it{Kepler}} Data}
\label{subsubsec:lckep}

In order to produce the best fit to the {\it{Kepler}} long cadence data, the model light curve is computed approximately every 10 minutes then binned to 29.4244 minute intervals using simple numerical integration. This ensures the model eclipse profiles are smoothed in a manner similar to the effective exposure time of {\it{Kepler}} long cadence data while maintaining feasible computation times.

Another consideration when using {\it{Kepler}} data is the aperture contamination from other stars due to the large pixels and apertures used. KIC~5738698 has several nearby stars that, though fainter, often contribute excess light. ELC attempts to account for this by using the contamination parameter (fraction of total flux contributed by nearby stars based on the final photometric aperture) reported by the Data Search database at MAST\footnote{\url{http://archive.stsci.edu/kepler/data_search/search.php}} to apply an offset to the model, given by $y_{\mathrm{off}} = (k*y_{\mathrm{med}})/(1-k)$, where $k$ is the value of the contamination parameter and $y_{\mathrm{med}}$ is the median value of the normalized flux light curve. The contamination parameter varies each quarter (between 0.010 and 0.021 for KIC~5738698) based on the orientation of the telescope, so we use the mean value of 0.015 when fitting the combined long cadence data. Quarter-to-quarter variations likely caused by the varying contamination values are discussed in \S\,\ref{subsubsec:kcont}.

 \subsubsection{Baseline Model}
 \label{subsubsec:lccirc}
\begin{figure}
\begin{center}
\includegraphics[scale=0.7,angle=90]{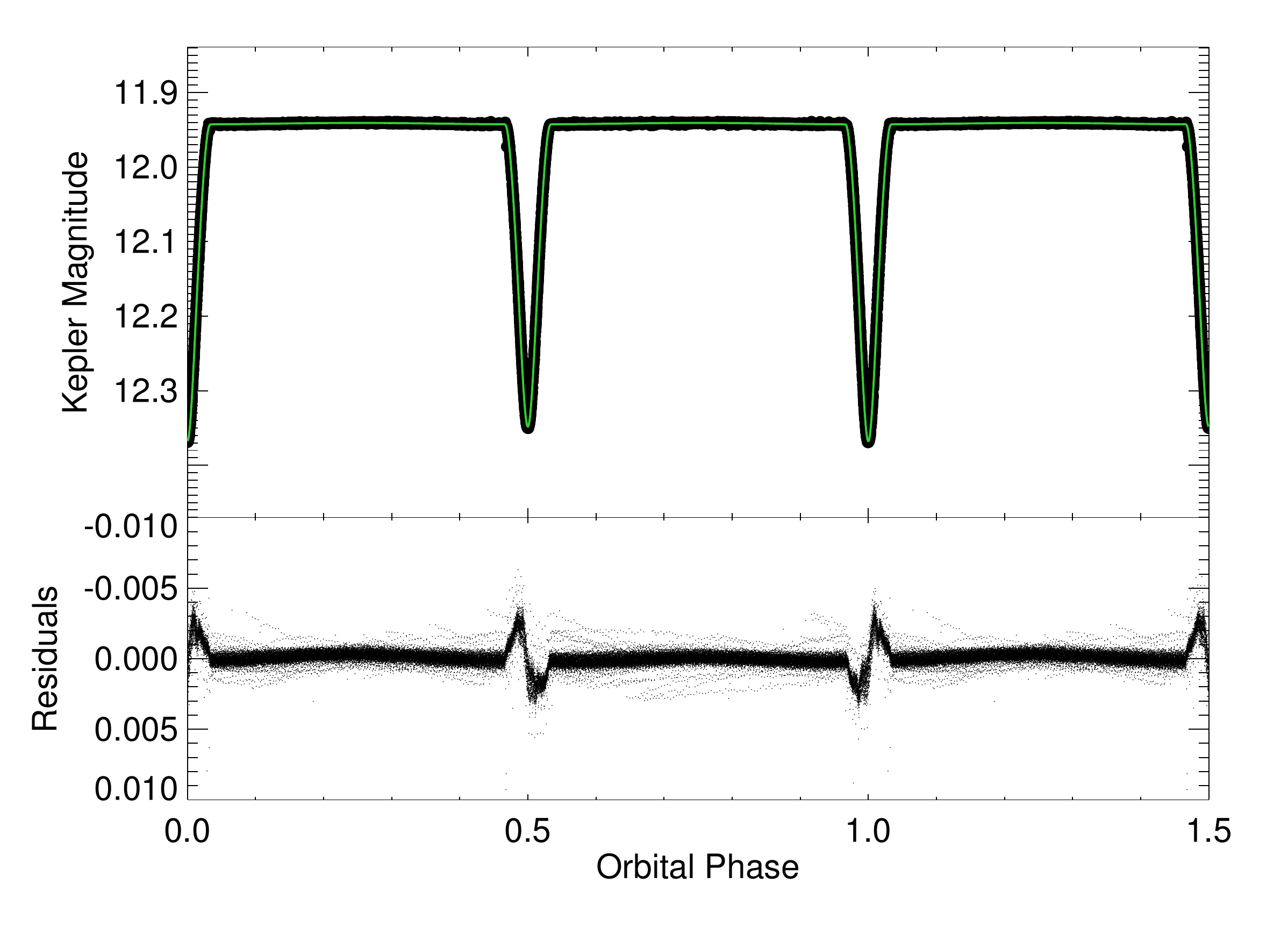}
\caption{Top: the phased, long cadence {\it{Kepler}} light curve of KIC~5738698 (black points) with the best-fit circular baseline model from ELC (solid green line). A randomly selected 20\% of the more than 65,000 data points are shown here. 
Bottom: residuals from the ELC fit to the {\it{Kepler}} light curve. See text for a discussion of the specific features and trends.}
\label{fig:lccirc}
\end{center}
\end{figure}

The best-fit model, as described above, for the long cadence light curve of KIC~5738698 is shown in Figure~\ref{fig:lccirc}. The {\it{Kepler}} photometry (black dots) and model (solid green line) are shown as well as the ($O - C$) residuals. Parameters used in the fit and their statistical uncertainties (see \S\ref{subsec:uncert}) are summarized in the first column of Table~\ref{tab:elc}. While the model approximates the overall light curve and eclipse shapes quite well, there is distinct structure in the residuals implying the fit is insufficient. In the next section we discuss several adjustments made to the model to minimize the residuals, including fitting for an eccentric orbit, examining the family of possible solutions over a range of inclinations and fractional radii, and considering the radiative properties of the stars.

\subsection{Improving the Fit/Minimizing Residuals}
\label{subsec:resid}

\subsubsection{In-Eclipse Residuals - Eccentric Solution}
\label{subsubsec:ecc}

Our circular baseline model for KIC~5738698 shows distinct features in the eclipse phase residuals, specifically a sine wave shaped component (see Fig.~\ref{fig:lccirc}), that indicate the positions and durations of the eclipses are not well fit. This implies that KIC~5738698 has a small, distinctly non-zero, eccentricity. In order to speed up convergence of an eccentric solution with ELC we derived initial estimates of the eccentricity~($e$) and longitude of periastron~($\omega$) from the light curve based on the offset between ($e \cos \omega$) and duration ($e \sin \omega$) of the two eclipses. The details of our approximations can be found in Appendices A (eclipse timings) and B (eclipse durations), but we summarize the process here. We began by determining the time of mid-eclipse for the primary and secondary by fitting a line through bisectors at different depths along each eclipse and extrapolating to the eclipse minimum. The phase difference between the primary and secondary is then used to determine $e \cos \omega$ according to the following relation (eq.~A12): 
\begin{equation}
e \cos \omega = \pi \frac{(\phi_{s} - \phi_{p} - 0.5)}{(1 + \csc i )}
\end{equation} 
where $\phi_{p},\phi_{s}$ are the phases of the primary and secondary eclipses, respectively, and the denominator ($1 + \csc i $) is used to approximate the effects of inclination when $i < 90\degr$ (see Appendix~A). For KIC~5738698 we find $e \cos \omega$ to be $0.00035,$ 
which is consistent with the value 0.000357 derived from our eclipse timing results \citep{Gies:aj2015a}.

We then measure eclipse durations ($d = t_{\rm{last}}- t_{\rm{first}}$) using a fit to the eclipse bisector widths extrapolated to the out-of-eclipse continuum. 
The ratio of the difference in eclipse times over their sum is then related to $e \sin \omega$ by (eq.~B5)
\begin{equation}
e \sin \omega = \frac{(d_{s} - d_{p})}{(d_{s} + d_{p})} 
\frac{1}{m}.
\end{equation} 
Here $d_{p}$ and $d_{s}$ are the durations of the primary and secondary eclipses, respectively, while $m$ represents the slope of the linear relation between the ratio of the difference and sum of the eclipse durations and $e \sin \omega$ when the eccentricity is small. We use a grid of $(R_1+R_2)/a$ and inclination $i$ values to derive the slope $m$ from linear fits to the ratio $(d_s-d_p)/(d_s+d_p) = m \times e \sin \omega$ (see Fig.~B1). This enables us to estimate $m$ from the values of $(R_1+R_2)/a$ and $i$ derived from our circular ELC model. We can then estimate $e \sin \omega$ from the observed ratio $(d_s-d_p)/(d_s+d_p)$ (see eqs.~B6 and B7 in Appendix B) divided by this value of $m$. For the circular model of KIC~5738698, $i = 86\fdg32$ and $(R_1+R_2)/a = 0.2128$, which gives us a value of $m = 0.884$. We then determine $e \sin \omega$ to be $0.0017$.

These approximations ($e \cos \omega = 0.00035$, $e \sin \omega = 0.0017$) provide starting values of $e = 0.0017$ and $\omega = 78^{\circ}$ 
that we use to define regions of parameter space to explore with ELC. Using the genetic optimizer, ELC converged to values of $e = 0.0006$ and $\omega \sim 50\degr$. However, because the durations of the primary and secondary eclipses are nearly identical and our measurement of the durations is only approximate, $e \sin \omega$ is not well constrained 
and slightly different estimations can lead to a range of $e \sin \omega$ values where $0\degr < \omega < 90\degr$ or $270\degr < \omega < 360\degr$. To account for this we also allowed $\omega$ to vary between 270 and 360$\degr$, which resulted in solutions where $e = 0.0004$ and $\omega \sim 300\degr$ with approximately the same chi-squared. Fitting for $e \cos \omega$ and $e \sin \omega$ directly led to $e \cos \omega = 0.00034$ and $e \sin \omega = -0.00122$, corresponding to $e= 0.0016$ and $\omega \sim 280\degr$ 
 but the chi-squared was higher than solutions fitting for $e$ and $\omega$ directly. Based on the slightly lower chi-squared and better fit to the light curve we adopt $e = 0.0006$ and $\omega = 52\degr$. The fitted and derived parameters from the eccentric model are listed in Table~\ref{tab:elc} and the model fit and residuals are shown in Figure~\ref{fig:finallc}. Given the ambiguity in the longitude of periastron and range of possible eccentricities for the system, however, it might be more appropriate to establish an upper limit 
  (as advocated for spectroscopic binaries by \citealt{Lucy:aap2013a}) of $e \lesssim 0.0017$ for KIC~5738698.
 

Given the precision of {\it{Kepler}} light curves and the delay in the secondary eclipse, another factor that has to be considered is the light travel time across the binary orbit. \citet{Bass:apj2012a} found approximately one-third of the delay in the secondary eclipse relative to phase 0.5 of KIC~6131659 was due to light travel time. We determine the delay using (eq.~A13)
\begin{equation}
\Delta t_{LT} = \frac{P K_2}{\pi c} (1-q)
\end{equation}
where $\Delta t_{LT}$ is the time difference between eclipses due to light travel time, $P$ is the period, $K_2$ is the velocity semi-amplitude of the secondary, $q$ is the mass ratio, and $c$ is the speed of light. 
We can then compare this to the relative times of the primary and secondary eclipses in an eccentric orbit ($\Delta t_{e}$) using\begin{equation}
\Delta t_{e} \approx \frac{2 P e}{\pi} \cos \omega
\end{equation}
(see eq.~A9), where $P$ is the period, $e$ is the eccentricity, and $\omega$ is the longitude of periastron of the binary. With the appropriate values from Table~\ref{tab:sporb} and our adopted eccentric ELC model, we calculate the light travel time delay in KIC~5738698 to be 1.5\,s with a total delay between the primary and secondary eclipses of 98.8\,s. 
Even if we consider the range of possible $e$ and $\omega$ values from different ELC models, the light travel time delay is at most 3\% of the total delay and we therefore do not attempt to make any corrections in our models, but are aware it may be a source of additional uncertainty in our measurements of $e$ and $\omega$.

\subsubsection{Out-of-Eclipse Residuals - Stellar shapes and sizes}
\label{subsubsec:size}

\begin{figure}
\begin{center}
\includegraphics[scale=0.6,angle=90]{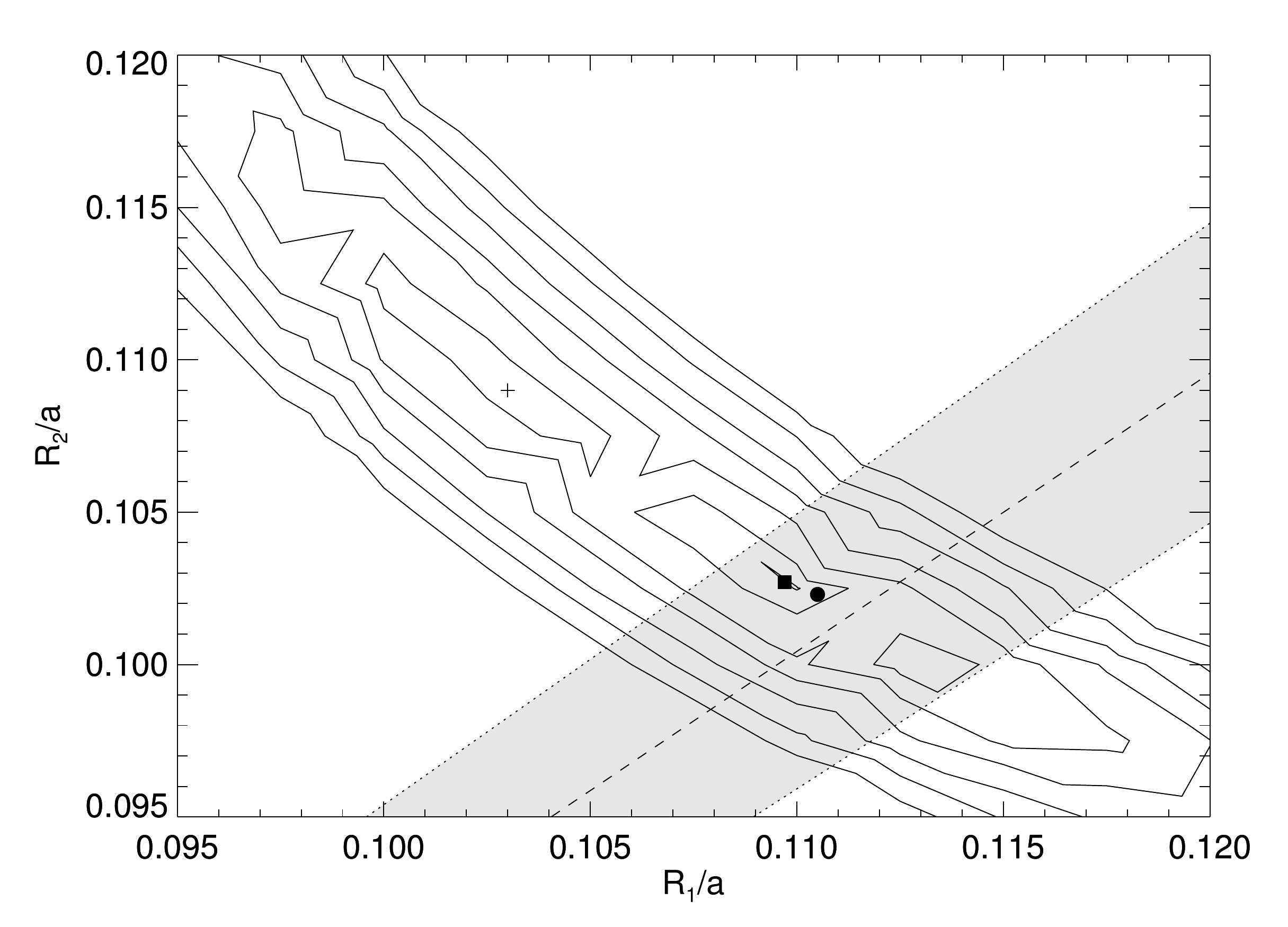}
\caption{Plot of chi-squared surface contours as a function of the fractional radii of the primary and secondary from a grid of values spanning $r_{1,2}$ = 0.095 - 0.12 with $\Delta r = 0.0025$. The contours represent regions 1 (behind square), 4, 9, 16, 25, 36, and 49 times the minimum chi-squared (arbitrarily chosen to highlight the topography), increasing outward from the valley through the center of the plot. The ratio of the radii, R$_{2}$/R$_{1}$ = 0.91$\pm$0.04, as derived from the spectroscopic flux ratio (\S4.4.2) is shown by the dashed line with the gray stripe representing the uncertainty. The filled circle and square show the fractional radii of our best-fit circular and eccentric ELC solutions, respectively. The plus sign gives the location of a solution where the primary star is smaller then the secondary, see text for details.}
\label{fig:contour}
\end{center}
\end{figure}

In addition to the signature of an eccentric orbit, the light curve residuals demonstrate a slight ellipsoidal variation outside of eclipse which was not well fit with the circular model. Figure \ref{fig:lccirc} shows that the residuals appear slightly brighter at quadrature relative to eclipse phases. We attempted to fit this modulation by adjusting the rotational distortion of the stars incrementally over a range of sizes and inclinations. To find combinations of fractional radii and inclination that would fit the observed light curve, we used a simple analytical model (employing linear limb darkening and {\it{Kepler}} contamination) to find values of $r_1$, $r_2$, and $i$ that were capable of reproducing the depths and widths of the primary eclipse. We then used the gridELC optimizer package, which uses a grid search routine to adjust a given set of parameters and find the minimum chi-squared, to find the model with the best fit to the observed light curve. The grid search produced two models with similar chi-squared values where the primary was larger in the first ($r_1$ = 0.110, $r_2$ = 0.101, $i$ = $86\fdg47$) and smaller in the second ($r_1$ = 0.103, $r_2$ = 0.109, $i$ = $86\fdg47$). We then produced model light curves using combinations of $r_1$ and $r_2$ spanning these values ($r_{1,2}$ = 0.095 - 0.12 with $\Delta r = 0.0025$) using our best-fit inclination ($i = 86\fdg33$, \S\ref{subsubsec:lcecc}). The chi-squared surface contours 
from these ELC models are plotted in Figure~\ref{fig:contour} as a function of the primary and secondary fractional radii. 

As demonstrated in the contour plot, for eclipsing binaries with partial, moderately deep, and nearly equal eclipses there exists a range of equally good fits due to a degeneracy among the inclination, radii, and secondary temperature \citep{Rozyczka:actaa2014a}. In order to remove the degeneracy and determine the best solution along the `valley' of fractional radii values, we use our spectroscopic flux ratio ($F_{2}/F_{1} = 0.82\pm0.06$; \S 3.3) and surface flux models from ATLAS9 
to calculate the ratio of the radii ($R_{2}/R_{1}$). The observed flux ratio is proportional to the projected areas and surface fluxes ($f_{2}/f_{1} = 0.98$) of the stars, such that 
\begin{equation}
\frac{R_{2}}{R_{1}} = \sqrt{\frac{F_{2}/F_{1}}{f_{2}/f_{1}}}.
\end{equation} 
Thus, the spectroscopic data impose the condition $R_{2}/R_{1}$ = 0.91$\pm$0.04, shown by the dashed line and gray region in Figure~\ref{fig:contour}. We therefore adopt solutions to the light curve in this region of the valley where the primary is larger than the secondary. The filled circle and square show the fractional radii of our best-fit circular (\S\,\ref{subsubsec:lccirc}) and eccentric ELC solutions (\S\,\ref{subsubsec:lcecc}), respectively. The plus sign gives the location of the minimum gridELC model where the primary star is smaller than the secondary for comparison.

While solutions in this region are mutually consistent with the light curve and spectroscopic data, the associated best-fit fractional radii do not account for the apparent ellipsoidal variation seen in the light curve.

\subsubsection{Out-of-Eclipse Residuals - Radiative Effects}
\label{subsubsec:alb}

In an effort to understand the out-of-eclipse modulation in our light curve residuals we next examined various radiative properties associated with binary modeling. As mentioned previously, we used the model atmospheres contained in ELC to compute surface intensities and thus account for the stellar limb darkening. In order to test whether this had any effect on the remaining residuals we produced ELC model light curves using linear and logarithmic limb darkening laws with coefficients from \citet{Howarth:mnras2011a}, \citet{Claret:aap2011a}, and the internal limb darkening tables (2011 version) of PHOEBE \citep{Prsa:apj2005a}. When using the ELC atmospheres to set the intensity at the surface normal and the logarithmic limb darkening law for all other angles, the eclipse depth is most affected. Similar changes in eclipse depth and very slight changes in eclipse widths occurred when local intensities were computed using a blackbody approximation with a linear or logarithmic limb darkening law. However, while the slight changes in eclipse depths and durations from the different parameterizations of limb darkening account for some of the modulation and scatter in the eclipse residuals, as expected they have no impact on the out-of-eclipse residuals.


Similarly, changing the gravity darkening exponents between the canonical value of 0.08 \citep{Lucy:zap1967a} for stars with convective envelopes and that derived from \citet{Claret:aap2000a} (see Table~\ref{tab:elc}) did not influence the residuals. 

As adjustments to the limb darkening and gravity brightening did not significantly improve the model fits to the out-of-eclipse variations, we included the bolometric albedo~$A$ of each star as fitted parameters in ELC. In the circular model of KIC~5738698 we fixed the albedo to 0.5, the canonical value for a star with a convective envelope. While \citet{Claret:mnras2001a} determined that the upper limit for convective envelopes should be 6300\,K, using 1.0 (the theoretical value of albedo for stars with radiative envelopes) greatly over estimated the out-of-eclipse flux. However, allowing the albedo to vary freely resulted in even lower values, with a best-fit solution of $A_{1}$ = 0.336 and $A_{2} = 0.334$. This suggests that the baseline model overestimated the flux around the eclipse times (when the illuminated hemispheres are directed our way), and thus the model residuals appeared somewhat fainter (Fig.~\ref{fig:lccirc}).

Using photometry from the WIRE satellite, \citet{Southworth:aap2007a} similarly found better fits to the amplitude of the light variation outside eclipse for the detached eclipsing binary $\beta$ Aurigae by including the bolometric albedos as fitting parameters. While the albedos they derived were similar to the theoretical value of 0.5 for convective atmospheres (0.59 and 0.56), the stellar temperatures (9350 and 9200\,K) suggest radiative atmospheres, indicating a lower then expected albedo. \citet{Southworth:mnras2011a} and \citet{Hambleton:ma2011a} also found unrealistic values ($>$1) of albedo when fitting eclipsing binaries using {\it{Kepler}} data.

While a lower albedo implies the stars re-radiate a smaller fraction of incident light then expected from theory, several past studies have found evidence for a broad range of albedo values. Initially \citet{Rucinski:actaa1969a} estimated the albedo for stars with convective envelopes to be between 0.4 and 0.5, and \citet{Rafert:mnras1980a} observationally determined a value slightly greater than 0.3 for two stars with temperatures similar to KIC~5738698. Furthermore, \citet{Sipahi:rmxaa2013a} derive $b$ and $v$-band albedos between 0.2 and 0.3 for three near contact binaries with convective secondaries ($T_{\rm{eff}}\sim 5500$\,K). This discrepancy between theory and observation revealed by the {\it{Kepler}} data serves to highlight our lack of understanding concerning the physical processes in the photospheres of stars near the transition zone between radiative and convective envelopes.

\subsubsection{Final LC Model}
\label{subsubsec:lcecc}

\begin{figure}
\begin{center}
\includegraphics[scale=0.7,angle=90]{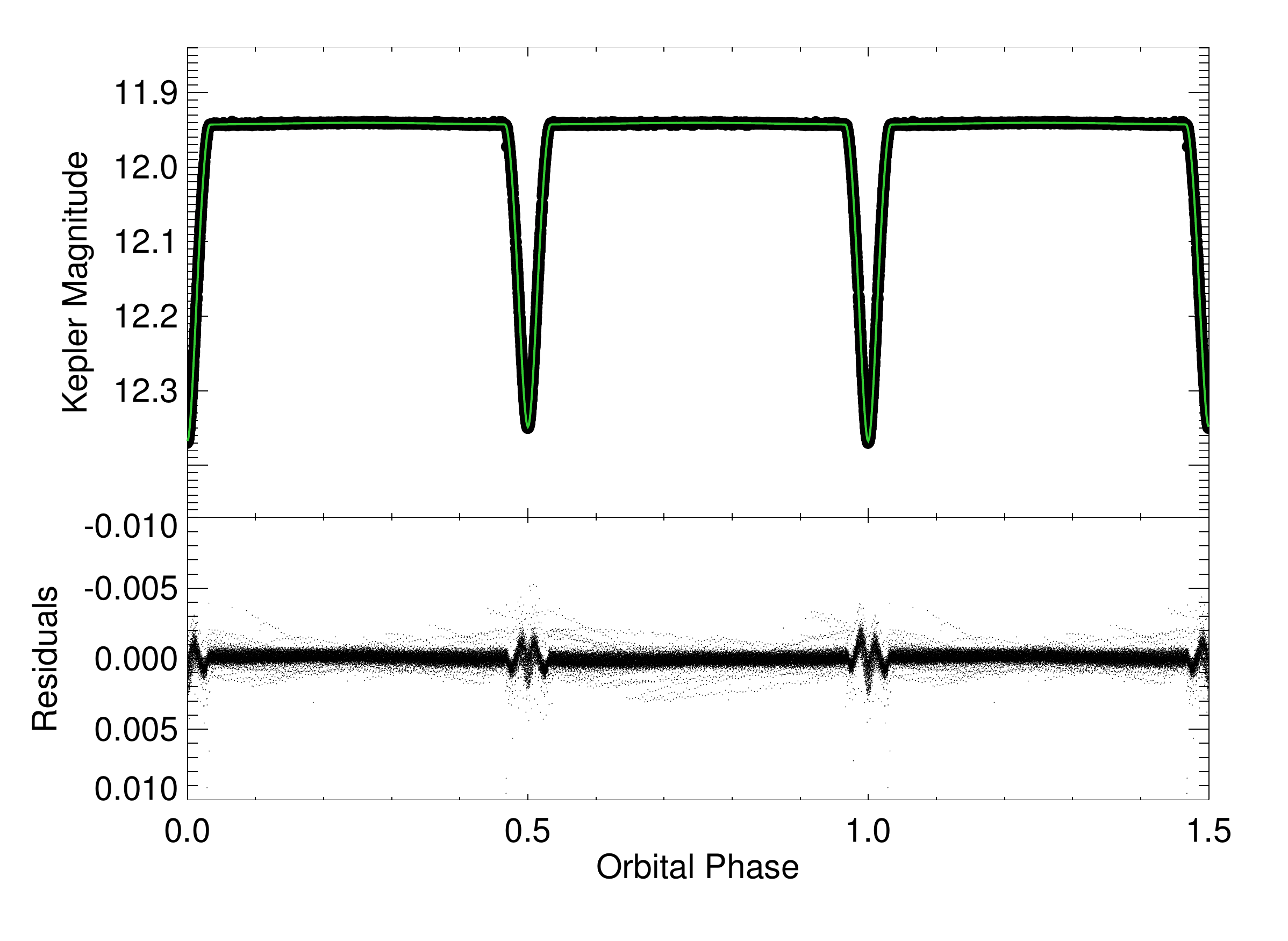}
\caption{Top: the phased, long cadence {\it{Kepler}} light curve of KIC~5738698 (black points) with the best-fit eccentric model from ELC (solid green line), as described in \S \ref{subsec:resid}. A randomly selected 20\% of the more than 65,000 data points are shown here. Phase zero is set as the time of primary eclipse. Bottom: residuals from the ELC fit to the {\it{Kepler}} light curve.}
\label{fig:finallc}
\end{center}
\end{figure}

Initial results from our baseline light curve model, updated radial velocity fit, and the values of $r_1$, $r_2$, $i$, $e$, and $\omega$ as derived above were then used as constraints for final models with ELC. The values of $i$, $r_1$, $r_2$, $T_2/T_1$, $e$, $\omega$, and $T_{\rm{0}}$ were allowed to vary and the genetic algorithm was used to determine the best-fit model. 
Our adopted ELC model light curve is shown in Figure~\ref{fig:finallc} with the {\it{Kepler}} data (black dots),  model (solid green line), and $(O-C)$ residuals shown at bottom. The fitted and calculated parameters are listed in Table~\ref{tab:elc}.

As can be seen in Figure~\ref{fig:finallc}, the residuals still show some structure, especially during eclipse. These likely reflect a combination of imperfect light curve normalization and quarterly changes in contamination values. 
In addition, \citet{Hambleton:mnras2013a} point out that small model discrepancies during light curve minima are common in the case of very accurate satellite light curves due to the incomplete physics in presently available models, while \citet{Sandquist:apj2013a} cite differences in the vertical structure of the ELC atmosphere models and observations as potential causes of systematic uncertainties in the derived parameters, which can be seen as mismatches between the model and observations.

\subsection{Parameter Uncertainties}
\label{subsec:uncert}

In order to estimate the statistical uncertainties on the fitted and derived astrophysical parameters we collapse the $n$-dimensional $\chi^2$ function from ELC onto each parameter of interest as in \citet{Orosz:apj2002a}. 
We scale the chi-squared values such that $\chi_{min}^2/\nu \approx 1$ and plot the lower envelope of each parameter by determining the minimum chi-squared in small bins across the whole range. 
The 1 and 2\,$\sigma$ confidence intervals are the parameter values where the lower envelope of the $\chi^2$ function is equal to $\chi_{min}^2 + 1$ and $\chi_{min}^2 + 4$, respectively.

\subsubsection{Long Cadence Segments and Short Cadence Data}
\label{subsubsec:segs}

As a check on systematic uncertainties in the light curve modeling and to examine any parameter variations over time, we divided our 18 quarters of long cadence data into eight individual segments. Each segment spanned two quarters (except the first and last groups which spanned three quarters; Q0-2 and Q15-17) with an average of $\sim$8000 data points. ELC models were run on each segment individually using the same input and fitting parameters as the `final' eccentric fit to the combined long cadence data. Table \ref{tab:segs} shows the free parameters for each segment as well as the mean and standard deviation. In each segment the longitude of periastron was allowed to vary between $0\degr < \omega < 90\degr$ and $270\degr < \omega < 360\degr$, resulting in two distinct clusters of $\omega$ as seen in Table \ref{tab:segs}. The values of $\omega$ oscillate in every other segment between mean values of $314\pm23\degr$ and $50\pm20\degr$. This highlights that while the tiny eccentricity is detectable in the light curve residuals, $\omega$ is not well constrained by the ELC models.

We also fit the single month of short cadence data for KIC~5738698 collected in Q4 ($\sim$45,000 data points). Because of the different `exposure times' of the long and short cadence data they cannot be modeled together, so we treated the short cadence data as a ninth segment and determined an optimized ELC model for the light curve. All of the inputs and model constraints were the same as those of the other segments except the model was not binned 
and the contamination parameter of 0.017 associated with the short cadence data was used. The parameters from the short cadence model are given in Table \ref{tab:segs} and agree very well with the other segments.

Because the spread of derived parameter values obtained in the segments should give us an independent measure of the systematic uncertainties, we adopt the standard deviations as our uncertainties in Table \ref{tab:elc}. 

\subsubsection{Aperture Contamination}
\label{subsubsec:kcont}

Our adopted value of the contamination factor as the fraction of flux from other stars in the {\it Kepler} pixel aperture presents another possible source of systematic uncertainty in our parameter estimates. Changes in the pixel aperture used in successive observing quarters can lead to changes in the contamination factor as faint nearby stars are included or excluded from the photometric summation over the aperture. We checked for this by forming phase binned light curves for each observing quarter and comparing the eclipse depths to those in a global average, phase binned light curve. We find that there is a small variation in eclipse depth that repeats over a four quarter cycle with a total amplitude of $0.5\%$ and which appears to correspond to maximum depth when the pixel aperture largely excludes three faint nearby stars (KIC~5738680, 5738689, and 5738720). The ratio of the total $V$-band flux of these three stars to the flux of KIC~5738698 is also $0.5\%$ according to the magnitudes reported by \citet{Everett:pasp2012a}, consistent with the idea that dilution of the eclipsing binary flux by these faint companions causes a very small change in eclipse depth. We think that the changes in the derived binary inclination 
with samples from different quarters (Table \ref{tab:segs}) is the result of the changing contamination factor. 


\subsection{Non-Orbital Frequencies}
\label{subsubsec:kcont}

\begin{figure}
\begin{center}
\includegraphics[scale=0.9,angle=0]{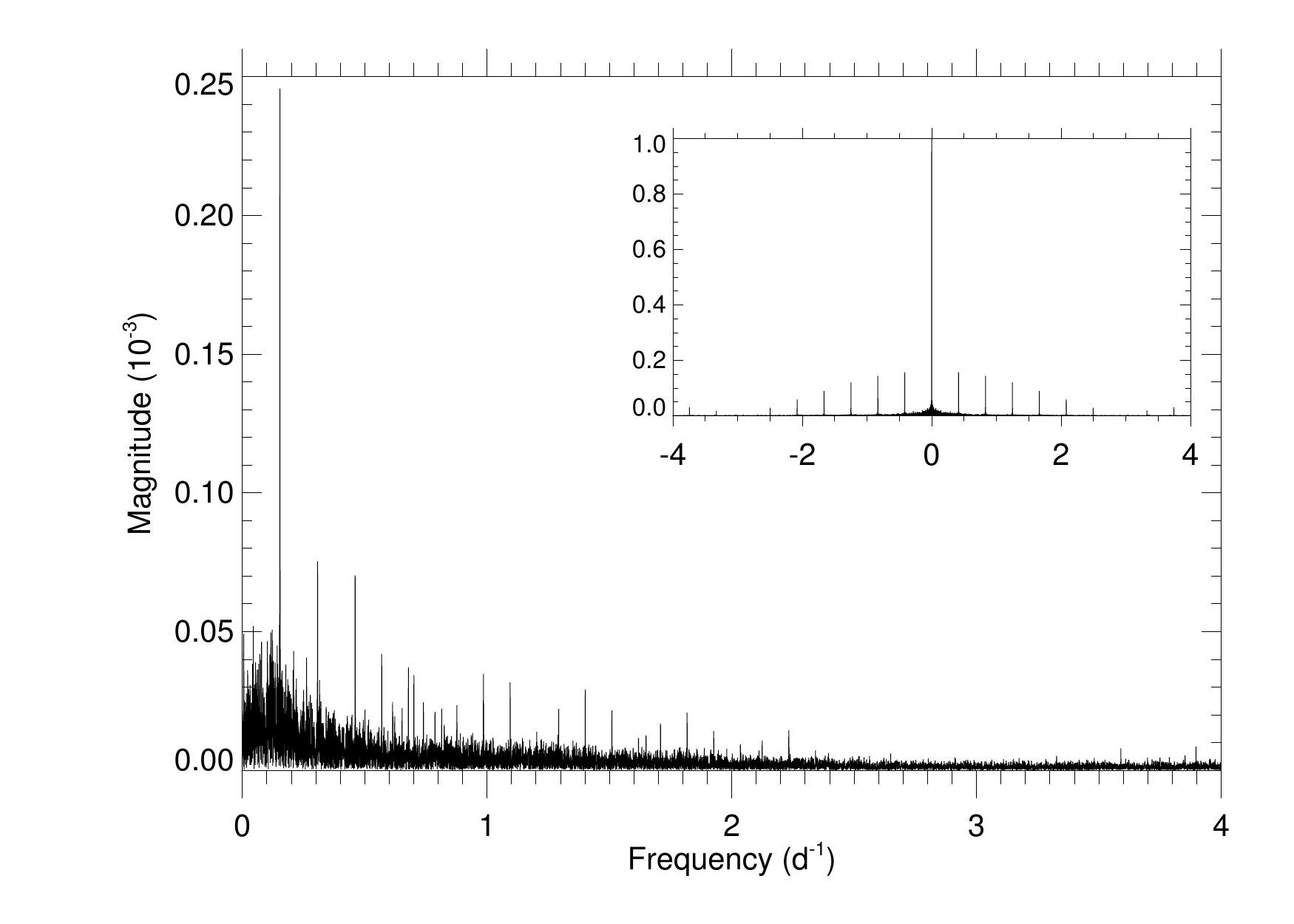}
\caption{The Fourier spectrum of the long cadence light curve residuals for KIC~5738698. The dominant frequencies are $f_1=0.15347$, $f_2=0.30725$, and $f_3=0.46067$ d$^{-1}$. The inset shows the spectral window function, which indicates the locations of the alias peaks introduced by removing the eclipse portion of the light curve resulting in 
gaps equal to twice the orbital frequency.}
\label{fig:puls}
\end{center}
\end{figure}

\citet{Gies:aj2012a} detected a hint of pulsation in the gray-scale diagram that depicts the difference from the mean light curve for each photometric measurement through Quarter 9 of KIC~5738698. We therefore calculated the Fourier transform of the long cadence residuals to detect any pulsational frequencies. The Fourier spectrum, shown in Figure \ref{fig:puls}, has a dominant peak at $f_1=0.15347\pm0.00002$ d$^{-1}$ (period of 6.51593 days) with two harmonics, $f_2=0.30725\pm0.00002$ d$^{-1}$ and $f_3=0.46067\pm0.00002$ d$^{-1}$. The observed frequencies appear similar to those generally found for high order-g modes and both components of KIC~5738698 do fall inside the $\gamma$ Doradus instability strip when placed on the $T_{\rm{eff}} - L$ plane \citep{Dupret:aap2005a}, making it possible one of the components is a $\gamma$ Doradus variable. We note, however, the period of $f_1$ is longer than the expected range for $\gamma$ Doradus stars \citep[$0.3-3$\,d;][]{Bradley:aj2015a} and it is also possible the frequency is due to rotational modulation. Nevertheless, the frequencies differ from the orbital period of the binary and should not affect our analysis of the light curve but will add to the residual scatter.




\section{Comparison with Evolutionary Models}
\label{sec:evol}

In the following, we compare the newly derived parameters of KIC~5738698 with those predicted by stellar evolutionary models. As masses and radii are the most directly testable parameters from eclipsing binaries, we compare our results with isochrones from a selection of evolutionary models in the $M-R$ plane through a chi-squared goodness-of-fit statistic. We determine the best fitting metallicity and age for our derived parameters by forming a grid of isochrones over a range of precomputed $\log Z/Z_\odot$ values and ages. The masses, radii, temperatures, and metallicities of the isochrones at each grid point are compared to our derived values using the following chi-squared criterion:
\begin{equation}
\chi^2 = \left[\sum_{i=1}^{2} \left(\frac{M-M_i}{\Delta M_i}\right)^{2}+\left(\frac{R-R_i}{\Delta R_i}\right)^{2}+\left(\frac{T-T_{i}}{\Delta T_{i}}\right)^{2}\right]+\left(\frac{\log \frac{Z_m}{Z_\odot}-\log \frac{Z_o}{Z_\odot}}{\Delta \log \frac{Z_o}{Z_\odot}}\right)^{2}
\end{equation}
where $M_{1,2}$, $R_{1,2}$, and $T_{1,2}$ are the masses, radii, and temperatures determined via ELC for the primary and secondary components (see Table \ref{tab:elc}); $M$, $R$, and $T$ are the corresponding values for the selected model isochrone; $Z_m$ is the metallicity of the selected isochrone; and $Z_o$ is the observed metallicity derived from spectroscopy ($-0.4\pm0.1$; \S \ref{subsec:tomog}). We find the minimum chi-squared for each $\log Z/Z_\odot$ and age, then use a spline fit to interpolate to the global minimum and determine the corresponding best-fit metallicity and age. To compare the metallicity with our spectroscopic results, we transform $\log Z/Z_\odot$ in terms of the solar metallicity used by each individual model (see Table \ref{tab:evolmod}) to the solar metallicity used in the UVBLUE spectral grids ($Z_{UV_\odot}=0.01886$). Uncertainties are found by using the degrees of freedom (5) to scale the chi-squared values and selecting the $1\sigma$ confidence interval where $\chi^2 = \chi_{min}^2+ 1$. The best-fit values and corresponding (unscaled) minimum chi-squared are given in the last three columns of Table \ref{tab:evolmod}. The unscaled minimum chi-squared for most of the models is well below the number of degrees of freedom indicating the parameter uncertainties are overestimated, however we use it as a means to intercompare the various models examined here.

\begin{figure}
\begin{center}
\includegraphics[scale=0.65,angle=0]{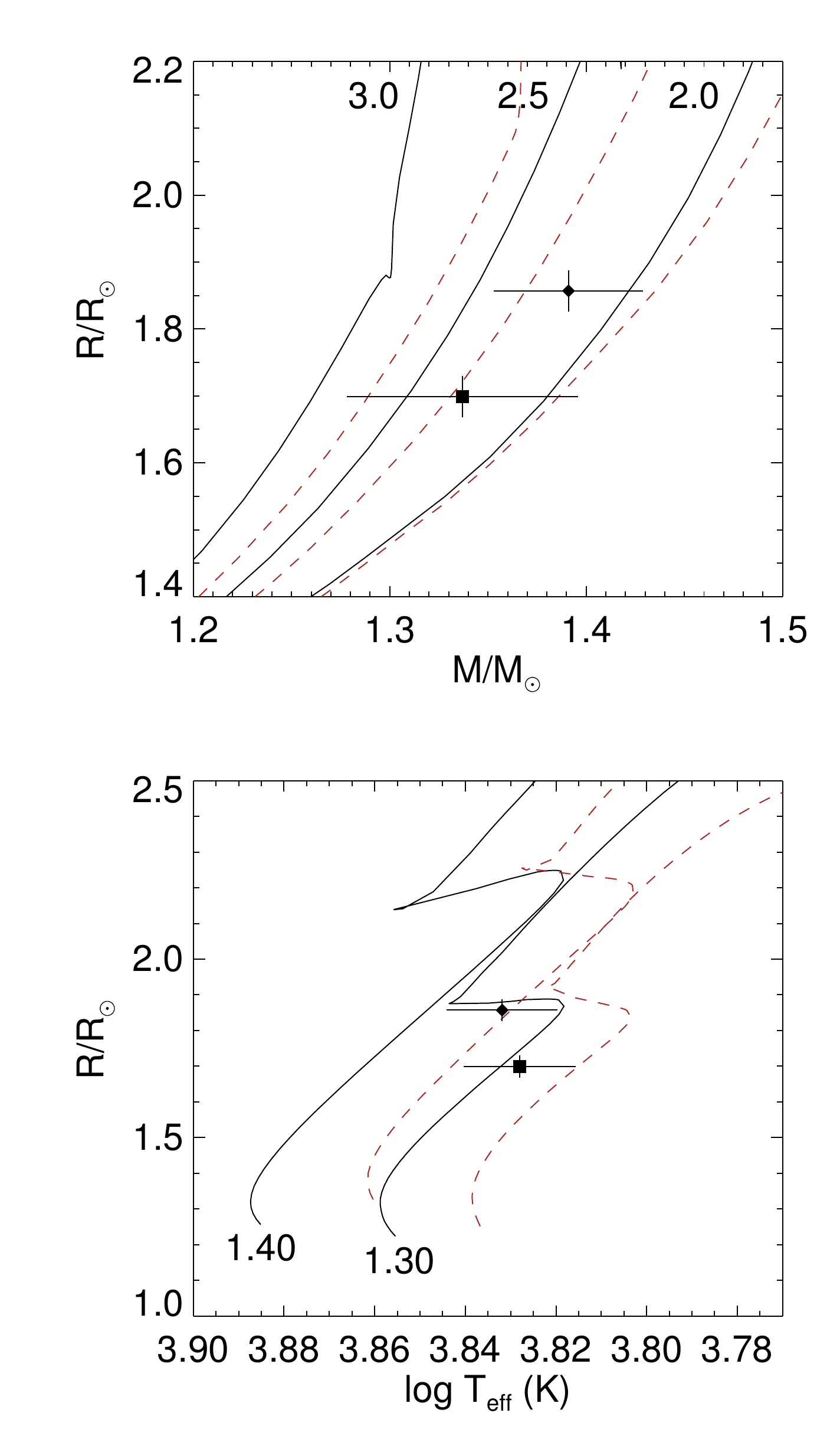}
\caption{Yonsei-Yale isochrones and evolutionary tracks plotted against the primary (diamond) and secondary (square) of KIC~5738698. Top: isochrones for $\log Z/Z_{UV_\odot} = -0.43$ (solid black lines) and $\log Z/Z_{UV_\odot} = -0.28$ (dashed red lines) with ages of (right to left) 2.0, 2.5, and 3.0~Gyr, as marked to the left of the isochrones. The metallicities have been scaled to the solar metal mass fraction used in the UVBLUE models, see \S\ref{sec:evol} for details. Bottom: evolutionary tracks for $\log Z/Z_{UV_\odot} = -0.43$ (solid black lines) and $\log Z/Z_{UV_\odot} = -0.28$ (dashed red lines) at (right to left) 1.3 and $1.4M_{\odot}$. Plots for each evolutionary model are available in the online version of the Journal (Figs. 8.1$-$8.4).} 
\label{fig:evolyy}
\end{center}
\end{figure}

The Yonsei-Yale (Y$^2$) isochrones\footnote{\url{http://www.astro.yale.edu/demarque/yystar.html}} by \citet{Demarque:apjs2004a} provide the best fit to the parameters of KIC~5738698. These models include an updated treatment of convective core overshoot in which the overshooting parameter $\Lambda_{OS}$ `ramps up' depending on the mass of the star and the metallicity, affecting the critical mass above which stars have a substantial convective core on the main sequence ($M_{\rm{crit}}^{\rm{conv}}$). More details of the input physics in the Y$^2$ models can be found in Table~\ref{tab:evolmod}. Isochrones and evolutionary tracks for $X = 0.749$ and $Z = 0.007$ (corresponding to $\log Z/Z_{UV_\odot} = -0.43$; solid black lines) and $X = 0.740$ and $Z = 0.010$ (corresponding to $\log Z/Z_{UV_\odot} = -0.28$; dashed red lines) are shown in Figure \ref{fig:evolyy}. While the uncertainties in the parameters allow for a range of possible ages, the best fit occurs at 2.32~Gyr with a metallicity of $\log Z/Z_{UV_\odot} = -0.31$. As seen in the $M-R$ plane, the isochrones have a steeper slope then that of a line connecting the primary (diamond) and secondary (square) suggesting the primary is younger. Similar age discrepancies have been observed in other F-type eclipsing binaries by \citet[][see their Figures~9 and 10]{Clausen:aap2010a} and \citet{Torres:aj2014a}, although they concentrate on systems with unequal masses in the range $1.15 - 1.70 M_\odot$. Both papers conclude that the age disparity in this particular mass range is likely due to the calibration of convective overshooting, though \citeauthor{Torres:aj2014a} suggest it may arise from a more complex relationship between overshooting, mass, and metallicity, possibly involving the evolutionary state as well. The evolutionary tracks plotted in the $T_{\rm{eff}}-R$ plane of Figure~\ref{fig:evolyy} show an offset between the derived masses and Y$^2$ theoretical tracks, such that both stars are undersized and/or cooler at our spectroscopically derived metallicity. When compared with the $\log Z/Z_{UV_\odot} = -0.28$ tracks (most similar to the best-fit metallicity of $-0.31$), however, both components fall along the corresponding mass track within the uncertainties. While this slightly higher metallicity appears to bring our observations in line with the models, we note that \citet{Clausen:aap2010a} similarly found the components of V1130 Tau ($1.31 M_{\odot}$ and $1.39 M_{\odot}$, $1.49 R_{\odot}$ and $1.78 R_{\odot}$, 6650\,K and 6625\,K, $P$ = 0.8\,d) approximately 200\,K cooler then the corresponding Y$^2$ models (see their Figure~6) at their observed metallicity of $\log Z = -0.24$ which may indicate a discrepancy with theory, although (like us) their abundances have not been derived in detail. 

We also compare our results to the Victoria-Regina Stellar Models\footnote{\url{http://www.cadc-ccda.hia-iha.nrc-cnrc.gc.ca/community/VictoriaReginaModels/}} from \citet{VandenBerg:apjs2006a}. These models determine the convective core boundary using integral equations for the maximum size of the central convective zone based on the luminosity from radiative processes and nuclear reactions using the free parameter $F_{\rm{over}}$, calibrated via open cluster color-mass diagrams. $F_{\rm{over}}$ is treated as a continuously increasing function between 1 and $2 M_{\odot}$, with the transition mass range adjusted for varying metallicities. In the $M-R$ plane 
(Fig.~8.2 in electronic version) 
the best-fit isochrones correspond to a younger age of 2.16~Gyr but nearly the same metallicity, $-0.30$, as the Y$^2$ models. In the $T_{\rm{eff}}-R$ plane the Victoria-Regina evolutionary tracks are very similar to the Y$^2$ models, although the $1.4M_\odot$ tracks turn off at slightly cooler temperatures and larger radii. 

Next we consider the PARSEC Stellar Evolution Code\footnote{\url{http://people.sissa.it/~sbressan/parsec.html}} of \citet{Bressan:mnras2012a} (v1.2s). Similar to the previously discussed models, PARSEC adopts a variable overshoot parameter that linearly increases throughout a transition region dependent on the metallicity. However, they define overshooting based on the mean free path of convective bubbles {\it{across}} the border of the convective region, with a maximum $\Lambda_{\rm{C}}$ of 0.5 which roughly coincides with $\Lambda_{\rm{OS}} = 0.25$ above the convective border as in other parameterizations \citep{Bressan:mnras2012a}. The best-fit isochrones in the $M-R$ plane 
(Fig.~8.3) 
indicate an age of 2.18~Gyr, consistent with the Victoria-Regina models, but less metal-poor with $\log Z/Z_{UV_\odot} = -0.23$. In the $T_{\rm{eff}}-R$ plane the tracks are slightly warmer then those of the Y$^2$ and Victoria-Regina models and the components of KIC~5738698 are further from the `blue hook' than in any of the other models.

Finally, we compare our derived parameters with the Geneva\footnote{\url{http://obswww.unige.ch/Recherche/evol/-Database-}} stellar evolution code of \citet{Mowlavi:aap2012a}. Here, the adopted overshoot treatment involves applying an overshoot parameter of $\Lambda_{\rm{OS}} = 0.10$ for $M > 1.7 M_{\odot}$, and half that between 1.25 and $1.7 M_{\odot}$. The chi-squared from the isochrones indicates an age of 2.26~Gyr, midway between the other age estimates, and $\log Z/Z_{UV_\odot} = -0.37$. This metallicity is the most metal-poor result from among the models examined here, but is the most consistent with our spectroscopic results. In the $T_{\rm{eff}}-R$ plane 
(Fig.~8.4), 
however, the Geneva models place the primary and secondary components very near to or even in the contraction phase of the `blue hook'. As this evolutionary stage is unlikely due to the short timescales involved, and overshooting results in extra hydrogen fuel in the core that lengthens the main sequence lifetimes of stars \citep{Lacy:aj2008a}, we speculate that the amount of overshooting applied in the Geneva models may be underestimated. For stars with our derived masses, the Geneva overshooting parameter is $\Lambda_{\rm{OS}} = 0.05$, compared to $\Lambda_{\rm{OS}} = 0.10$ at $1.3 M_{\odot}$ and $\Lambda_{\rm{OS}} = 0.15$ at $1.4 M_{\odot}$ in the Y$^2$ models.

\section{Summary}
\label{sec:sum}

We have analyzed $\sim$3.5 years of {\it{Kepler}} photometric data along with supporting ground-based optical spectra to solve for the orbital and physical properties of the eclipsing binary KIC 5738698. Through radial velocity measurements and reconstruction of the individual spectra we find effective temperatures of 6790 and 6740\,K for the primary and secondary, respectively. In modeling the light curve we have highlighted the detail probed by {\it{Kepler}} which allows us to consider a tiny orbital eccentricity and the lower than expected value of albedo. The parameters derived from the radial velocity and light curves indicate the binary consists of two very similar stars ($1.39M_\odot$, $1.34M_\odot$; $1.84R_\odot$, $1.72R_\odot$). Comparisons with stellar evolutionary models suggest the components are slightly less metal-poor then we estimated from spectroscopy, though still within our uncertainty. This minor discrepancy may indicate we underestimated the flux contamination from nearby stars, as extra light in the spectra would result in weaker lines that can mimic the effects of a smaller $\log Z/Z_\odot$. At this time, however, we find the best agreement with the Y$^2$ models for $\log Z/Z_\odot = -0.31$ and an age of 2.3~Gyr. 


This work exploits the precise photometry and long time baseline of {\it{Kepler}} to add to the known eclipsing binaries with accurate masses and radii (within 4\% and 2\%, respectively), adding to a small sample of stars located at the end of the core hydrogen burning phase which are sensitive to the amount of convective overshooting adopted in models. However, further benefit would come from high resolution spectra of KIC~5738698 in order to derive detailed abundances that could either reinforce the abundances suggested by the evolutionary models or indicate the presence of a companion. 
Such spectra would also result in tighter constraints on the effective temperatures and masses (through more RV measurements) of the stellar components.

Future articles in the series will contain similar analyses of other eclipsing binaries in our eclipse timing and spectroscopic studies. We are also conducting pulsational analyses of systems showing $\delta$ Scuti/$\gamma$ Doradus pulsations \citep[e.g.][ApJ, \it{submitted}]{Guo:apj2016a}.

\acknowledgments

We acknowledge the observations taken using the 4\,m Mayall telescope at KPNO and are grateful to the director and staff of KPNO for their help in obtaining these observations. We also thank the anonymous referee for providing constructive comments that improved the paper. {\it Kepler} was competitively selected as the tenth Discovery mission. Funding for this mission is provided by NASA's Science Mission Directorate. This study was supported by NASA grants NNX12AC81G, NNX13AC21G, and NNX13AC20G.
This material is based upon work supported by the National Science Foundation under Grant No.~AST-1009080 and AST-1411654. Institutional support has been provided from the GSU College of Arts and Sciences and from the Research Program Enhancement
fund of the Board of Regents of the University System of Georgia, administered through the GSU Office of the Vice President for Research and Economic Development.

{\it Facilities:} \facility{Kepler}, \facility{Mayall}, \facility{Perkins}.

\clearpage

\renewcommand\appendixname{Appendix}
\appendix

\section{Eclipse Timings and $e\cos\omega$}        
 
The time between primary and secondary eclipse will generally differ from half the orbital period $P$ for binary systems with a non-zero eccentricity. \citet{Kopal:1959}, \citet{Binnendijk:1960}, and \citet{Hilditch:2001} present an analytical solution for this difference for the case of inclination $i=90^\circ$ that we summarize here. Figure~A1 illustrates the orbital geometry for the elliptical orbit of the primary star.  This star orbits the center of mass attaining periastron at the right hand side of the diagram. Suppose we observe the binary along a line of sight from the lower left, so that primary eclipse (primary superior conjunction) occurs when the primary is at the location marked by a square and secondary eclipse occurs at the point indicated by a diamond. The true anomaly $\nu$ measures the angle from periastron to the eclipse position, and given a longitude of periastron for the primary $\omega$, then $\nu= \pi /2 -\omega$ at primary eclipse and $\nu= 3 \pi /2 -\omega$ at secondary eclipse. The relation between orbital phase and position is given by the Kepler equation 
\begin{equation}
{{2\pi (t-T)}\over{P}} = E - e \sin E 
\end{equation}
where $T$ is the epoch of periastron, $e$ is the eccentricity, and $E$ is the eccentric anomaly. The angle $E$ is measured from periastron through the center of the ellipse to the 
stellar position projected onto the auxiliary circle. Then the time between eclipses is given by
\begin{equation}
{{2\pi (t_s-t_p)}\over{P}} = E_s - E_p - e (\sin E_s - \sin E_p) 
\end{equation}
where $E_p$ and $E_s$ are the values of the eccentric anomaly at the primary and secondary eclipses, respectively. 


The true and eccentric anomalies are related through the expression for the ratio of binary separation $r$ to semimajor axis $a$,
\begin{equation}
{{r}\over{a}} = {{(1-e^2)}\over{(1+e\cos \nu)}} = 1 - e \cos E. 
\end{equation}
We can determine $\cos E_p$ from the above as 
\begin{equation}
\cos E_p = (1 - {{r_p}\over{a}})/e 
\end{equation}
where $r_p$ is the center of mass to primary distance at primary eclipse (and similarly for the secondary eclipse). From inspection of Figure~A1, we can derive an expression for $\sin E_p$, 
\begin{equation}
\sin E_p = {{r_p}\over{a}} {{\sin \nu }\over{b/a}} = {{r_p}\over{a}} {{\cos \omega }\over{b/a}}
\end{equation}
where the ratio of minor to major axis is $b/a=(1-e^2)^{1/2}$. The comparable expression for the secondary eclipse is 
\begin{equation}
\sin E_s = -{{r_s}\over{a}} {{\cos \omega }\over{b/a}}. 
\end{equation}
We need to solve for the angle $2x \equiv y \equiv E_s - E_p - \pi$ shown in Figure~A1. The analytical solution is obtained by using the identity relation for the tangent of a half angle 
\begin{equation}
\tan {{y}\over{2}} = \sqrt{ {{1-\cos y}\over{1+\cos y}} } \\
 = \sqrt{{1+\cos E_s \cos E_p+\sin E_s \sin E_p}\over{1-\cos E_s \cos E_p-\sin E_s \sin E_p}}
 = {{e \cos\omega}\over{(1-e^2)^{1/2}}}.  
\end{equation}
Then we arrive at the final expression for the observed phase difference between eclipses, 
\begin{equation}
{{2\pi (t_s-t_p)}\over{P}} = \pi +2 \arctan {{e \cos\omega}\over{(1-e^2)^{1/2}}} 
 + {{2 (1-e^2)^{1/2} ~e \cos \omega }\over{(1 - e^2 \sin^2 \omega})}. 
\end{equation}

A Taylor series expansion for small $e$ of the right hand side of equation A8 yields 
\begin{equation}
{{2\pi (t_s-t_p)}\over{P}} = \pi + 4 e\cos\omega  
\end{equation}
or in terms of the orbital phase difference
\begin{equation}
e\cos\omega = {{\pi}\over{2}} (\phi_s - \phi_p -0.5). 
\end{equation}
This approximation for $e\cos\omega$ differs from the actual value by less than $10^{-3.8}$, $10^{-6.8}$, and $10^{-9.7}$ for $e = 0.1$, 0.01, and 0.001, respectively, so use of this approximation will introduce negligible errors in cases of small eccentricity and $i=90^\circ$.

The relationship will change somewhat for $i<90^\circ$. The projected separation in the sky $\delta$ in units of the semimajor axis will vary as \citep[see eq.\ 9 from][]{Gimenez:aap2006a}
\begin{equation}  
\delta^2 =\left[{{1-e^2}\over{1-e\sin (\theta - \omega )}}\right]^2
   (1 - \cos^2 \theta \sin^2 i)
\end{equation}
where $\theta = \nu + \omega - \pi /2$.  The central times of the eclipses correspond to the two minima of this function. 
We made a numerical solution to the eclipse time difference as a function of inclination, eccentricity, and longitude of periastron. Once again we find that the eclipse time difference is mainly related to the parameter $e\cos \omega$. However, direct use of equation A10 may lead to small overestimates of $e\cos \omega$ for $i<90^\circ$, and a good approximation for the effects of inclination (better than $0.5\%$ for $i>60^\circ$) is given by
\begin{equation}
e\cos\omega = {{\pi}}~ {{\phi_s - \phi_p - 0.5}\over{1 + \csc i}}. 
\end{equation}

Finally, we note that light travel time differences between eclipses may also cause a small difference in eclipse times given by \citep{Kaplan:apjl2010a, Fabrycky:2010a} 
\begin{equation}
\triangle t = t_s-t_p -{P\over 2}= {{P K_2}\over{\pi c}} \left(1 - {M_2\over M_1}\right) 
 {{(1-e^2)^{3/2}}\over{1-e^2\sin^2\omega}}
\end{equation}
where $K_2$ is the orbital semiamplitude of the secondary, $c$ is the speed of light, and $M_2 / M_1$ is the mass ratio. The last factor relating to the eccentricity is of order $1 - e^2$ 
and can be ignored for the small $e$ case. If radial velocity solutions are available for both components and $\triangle t$ is found to be significant, then the numerator factor in equation A12 should be replaced with $\phi_s - \phi_p -0.5 - \triangle t /P$ in order to estimate $e \cos \omega$. 

 
\section{Eclipse Durations and $e\sin\omega$}        

\citet{Kopal:1959}, \citet{Binnendijk:1960}, and \citet{Hilditch:2001} 
discuss how the eclipse durations are closely related to the product $e\sin\omega$. 
Figure~B1 shows the geometry for the eclipse as viewed in the sky. Suppose that first contact occurs on the left side when the stellar limbs first meet and that final contact occurs on the 
right side as the limbs last coincide. In the frame of reference of the primary star, the secondary moves a distance $2x$ where $x^2 = (R_p +R_s)^2 - \delta^2$ and $\delta = r \cos i$ (minimum separation). The relative projected velocity of secondary at the eclipse time is $v = {{2 \pi a}\over{P}} {{a}\over{r}}$ according to Kepler's Second Law (ignoring minor non-tangential motions). Then using the relations for the separation $r/a$ at both eclipses, the duration of the primary and secondary eclipses are, respectively,
\begin{equation}
d_p = {{P}\over{\pi}} {{1-e^2}\over{1+e\sin\omega}} 
      \sqrt{ \left({{R_p+R_s}\over{a}}\right)^2 - \left({{1-e^2}\over{1+e\sin\omega}}\right)^2 \cos^2 i}
\end{equation}
and 
\begin{equation}
d_s = {{P}\over{\pi}} {{1-e^2}\over{1-e\sin\omega}} 
      \sqrt{ \left({{R_p+R_s}\over{a}}\right)^2 - \left({{1-e^2}\over{1-e\sin\omega}}\right)^2 \cos^2 i}.
\end{equation}
Note that if eclipses do occur, then the arguments of the square root will be positively valued.  


It is helpful to consider a specific case to understand the dependencies in the above equations.  Suppose $\omega = 90^\circ$ so that primary eclipse occurs when the primary reaches periastron at its superior conjunction.  At this instance the separation is a minimum $(1-e)a$, so the Keplerian velocity reaches a maximum to cause a shorter duration eclipse.  This part of the variation is given by the leading term $(1-e^2)/(1+e\sin\omega)=(1-e)$ in the 
expression for $d_p$. However, if the eclipse is viewed with an inclination different from $i=90^\circ$, then the reduced separation between the stars will result in a proportional decrease in the 
impact parameter $\delta$ (Fig.\ B1) and hence an increase in the eclipse crossing distance $2x$. This change appears in the square root term in the expressions above.  Thus, the change in eclipse duration depends on the relative sizes of the competing terms of changing velocity (shorter eclipse in this case) and eclipse path length (longer eclipse in this case).  While both terms depend on $e\sin\omega$, the latter one also depends upon the sum of the stellar radii and inclination. 

We may then form an expression for the ratio of the difference in eclipse durations over their sum, 
\begin{equation}
{{d_s-d_p}\over{d_s+d_p}} = 
{
 {(1+e\sin\omega) \sqrt{1-\left({{1-e^2}\over{1-e\sin\omega}}\right)^2 \epsilon^2}
 -(1-e\sin\omega) \sqrt{1-\left({{1-e^2}\over{1+e\sin\omega}}\right)^2 \epsilon^2} }
 \over
 {(1+e\sin\omega) \sqrt{1-\left({{1-e^2}\over{1-e\sin\omega}}\right)^2 \epsilon^2}
 +(1-e\sin\omega) \sqrt{1-\left({{1-e^2}\over{1+e\sin\omega}}\right)^2 \epsilon^2} }
 }
\end{equation}
where $\epsilon = {{a \cos i}\over{R_p+R_s}}$.  For most eclipsing systems, $i\approx 90^\circ$ and $\epsilon^2 << 1$, and, therefore, we may use the Taylor series expansion for the square root terms, $\sqrt{1-x^2}\approx 1 -x^2/2$.  Then, ignoring terms of $e^2$ and higher order, the above ratio simplifies to
\begin{equation}
{{d_s-d_p}\over{d_s+d_p}} = {{2-3\epsilon^2}\over{2-\epsilon^2}} ~e\sin\omega 
 \equiv m ~e\sin\omega 
\end{equation}
which yields the estimator 
\begin{equation}
e\sin\omega = {{(d_s-d_p)}\over{(d_s+d_p)}} {{1}\over{m}}.
\end{equation}

It is usually possible to make a preliminary fit of an eclipsing light curve assuming a circular orbit for small eccentricity systems, and $m(\epsilon)$ may be estimated from the inclination and 
fractional radius sum associated with the circular fit.   

The simple rectilinear approximation above does not account fully for the three dimensional projection of the orbit during the eclipse, and a better approximation of the eclipse duration is given by \citet[][see his eq.\ 15]{Kipping:mnras2010a} 
\begin{equation}
d_p = {{P}\over{\pi}} {{(1-e^2)^{3/2}}\over{(1+e\sin\omega)^2}}
 \arcsin\left[ 
 {
 {\sqrt{ \left({{R_p+R_s}\over{a}}\right)^2 - \left({{1-e^2}\over{1+e\sin\omega}}\right)^2 \cos^2 i} }
 \over
 {\left({{1-e^2}\over{1+e\sin\omega}}\right) \sin i}
 }   
 \right]
\end{equation}
and 
\begin{equation}
d_s = {{P}\over{\pi}} {{(1-e^2)^{3/2}}\over{(1-e\sin\omega)^2}}
 \arcsin\left[ 
 {
 {\sqrt{ \left({{R_p+R_s}\over{a}}\right)^2 - \left({{1-e^2}\over{1-e\sin\omega}}\right)^2 \cos^2 i} }
 \over
 {\left({{1-e^2}\over{1-e\sin\omega}}\right) \sin i}
 }   
 \right].
\end{equation}
This approximation assumes that the separation at mideclipse is constant throughout the eclipse. In the limit when $i\approx 90^\circ$ and the sum of the radii is small, these expressions attain the same form as in equations B1 and B2.  However, equations B6 and B7 show more clearly that eclipse durations are really functions of two variables, the inclination and the sum of the relative radii, rather than a ratio of these parameters.  

The ratio of the difference and sum of equations B6 and B7 forms an expression like equation B5 that is also linear in $e \sin \omega$ for small eccentricity.  We show in Figure~B2 the derived slope $m$ from linear fits of the ratio $(d_s - d_p)/(d_s+d_p) = m \times (e\sin\omega)$ for a grid of summed radii $(R_1 + R_2 )/a$ and inclinations $i$. This graph can be used to estimate $m$ from the preliminary values of $(R_1 + R_2 )/a$ and $i$ derived using a circular fit to the light curve. Then $e \sin \omega$ may be estimated from the observed ratio $(d_s - d_p)/(d_s+d_p)$ divided by the slope $m$ from Figure~B2. Note that as the radius sum and inclination decline the  slope changes from a positive to negative sign as the relative sizes of the terms for changes in velocity and eclipse path length reverse. Thus, at some point along a constant $i$ curve there occurs $m=0$, meaning that small changes in $e \sin \omega$ result in no change in eclipse duration. In such a situation $e \sin \omega$ cannot be determined from the equal eclipse durations. We also show as a thin dashed line in Figure~B2 the estimate of $m$ derived from the first order expression (eq.\ B4) for the case of $i=85^\circ$. Comparing this to the more accurate calculation (solid line indicated by $i=85^\circ$), we see that the first order expression is only 
adequate over a small range in $(R_1 + R_2 )/a$ and can even have the wrong sign in some cases (small $(R_1 + R_2 )/a$).  Consequently, we recommend the use in practice of the calculations for $m$ shown in Figure~B2 rather than the approximation given in equation B4. 


The electronic version of the paper includes the IDL code {\tt findecc.pro} that may be used to determine estimates of $e \cos \omega$ and $e \sin \omega$ as described above.  A subroutine {\tt tkmod.pro} is included therein that calculates the slope term $m$ as a function of $(R_1 + R_2 )/a$ and $i$ using equations B6 and B7.  Our method uses a folded magnitude light curve that is binned in orbital phase. Each eclipse is identified and rescaled into a form with magnitude set to zero in the out-of-eclipse region and one at maximum eclipse depth.  Then the eclipse curve is subdivided into 18 depth points at $5\%$ intervals from 0.05 to 0.9, and the eclipse bisector is determined at each depth point.  The bisector midpoints are extrapolated to the eclipse core to estimate the phase of the eclipse, and the bisector widths are extrapolated to the out-of-eclipse level to find the duration of the eclipse.  Although this estimate of duration is an 
underestimate of the actual duration, it is sufficient for use in the difference over sum ratio to find $e \sin \omega$. We caution that the scheme assumes that the stars are spherical, ignoring the tidal distortions that are often present in close binaries.  Nevertheless, the method offers useful starting estimates for $e$ and $\omega$ for use in light curve modeling programs like ELC.

\clearpage

\bibliography{573ms_ArXiv.bbl}

\clearpage



\clearpage


\clearpage


\clearpage


\clearpage


\clearpage


\clearpage


\clearpage


\clearpage


\figsetstart
\figsetnum{8}
\figsettitle{Comparison with Evolutionary Models}


\figsetgrpstart
\figsetgrpnum{8.1}
\figsetgrptitle{Yonsei-Yale Isochrones}
\figsetplot{f8.eps}
\figsetgrpnote{Yonsei-Yale isochrones and evolutionary tracks plotted against the primary (diamond) and secondary (square) of KIC~5738698. Top: isochrones for $\log Z/Z_{UV_\odot} = -0.43$ (solid black lines) and $\log Z/Z_{UV_\odot} = -0.28$ (dashed red lines) with ages of (right to left) 2.0, 2.5, and 3.0~Gyr, as marked to the left of the isochrones. The metallicities have been scaled to the solar metal mass fraction used in the UVBLUE models, see \S\ref{sec:evol} for details. Bottom: evolutionary tracks for $\log Z/Z_{UV_\odot} = -0.43$ (solid black lines) and $\log Z/Z_{UV_\odot} = -0.28$ (dashed red lines) at (right to left) 1.3 and $1.4M_{\odot}$.}
\figsetgrpend






\figsetgrpstart
\figsetgrpnum{8.2}
\figsetgrptitle{Victoria-Regina Isochrones}
\figsetplot{f82.eps}
\figsetgrpnote{Victoria-Regina isochrones and evolutionary tracks plotted against the primary (diamond) and secondary (square) of KIC~5738698. Top: isochrones for $\log Z/Z_{UV_\odot} = -0.37$ (solid black lines) and $\log Z/Z_{UV_\odot} = -0.27$ (dashed red lines) with ages of (right to left) 2.0, 2.4, and 3.0~Gyr, as marked to the left of the isochrones. The metallicities have been scaled to the solar metal mass fraction used in the UVBLUE models, see \S\ref{sec:evol} for details. Bottom: evolutionary tracks for $\log Z/Z_{UV_\odot} = -0.37$ (solid black lines) and $\log Z/Z_{UV_\odot} = -0.27$ (dashed red lines) at (right to left) 1.3 and $1.4M_{\odot}$.} 
\label{fig:evolvr}
\figsetgrpend





\figsetgrpstart
\figsetgrpnum{8.3}
\figsetgrptitle{PARSEC Isochrones}
\figsetplot{f83.eps}
\figsetgrpnote{PARSEC isochrones and evolutionary tracks plotted against the primary (diamond) and secondary (square) of KIC~5738698. Top: isochrones for $\log Z/Z_{UV_\odot} = -0.37$ (solid black lines) and $\log Z/Z_{UV_\odot} = -0.28$ (dashed red lines) with ages of (right to left) 2.0, 2.5, and 3.0~Gyr, as marked to the left of the isochrones. The metallicities have been scaled to the solar metal mass fraction used in the UVBLUE models, see \S\ref{sec:evol} for details. Bottom: evolutionary tracks for $\log Z/Z_{UV_\odot} = -0.37$ (solid black lines) and $\log Z/Z_{UV_\odot} = -0.28$ (dashed red lines) at (right to left) 1.3 and $1.4M_{\odot}$.} 
\label{fig:evolp}
\figsetgrpend




\figsetgrpstart
\figsetgrpnum{8.4}
\figsetgrptitle{Geneva Isochrones}
\figsetplot{f84.eps}
\figsetgrpnote{Geneva isochrones and evolutionary tracks plotted against the primary (diamond) and secondary (square) of KIC~5738698. Top: isochrones for $\log Z/Z_{UV_\odot} = -0.50$ (solid black lines) and $\log Z/Z_{UV_\odot} = -0.28$ (dashed red lines) with ages of (right to left) 2.0, 2.5, and 3.1~Gyr, as marked to the left of the isochrones. The metallicities have been scaled to the solar metal mass fraction used in the UVBLUE models, see \S\ref{sec:evol} for details. Bottom: evolutionary tracks for $\log Z/Z_{UV_\odot} = -0.50$ (solid black lines) and $\log Z/Z_{UV_\odot} = -0.28$ (dashed red lines) at (right to left) 1.3 and $1.4M_{\odot}$.} 
\label{fig:evolg}
\figsetgrpend




\setcounter{figure}{0}
\renewcommand{\thefigure}{A\arabic{figure}}

\begin{figure}
\begin{center} 
{\includegraphics[angle=0,height=12cm]{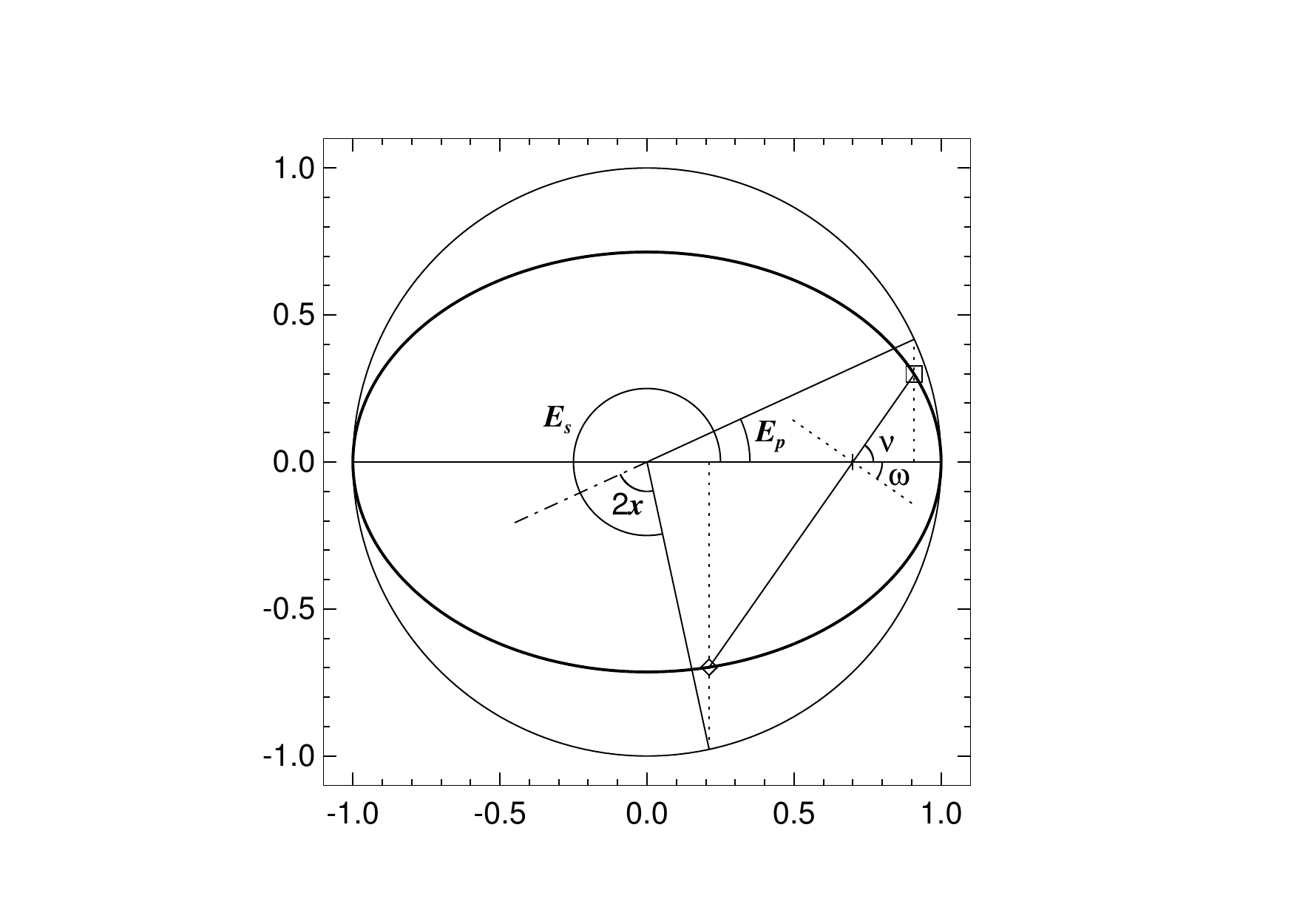}}
\end{center}
\caption{A diagram of an elliptical orbit and eclipse geometry as seen from above the orbit. The thick solid ellipse shows the elliptical orbit of the primary star with a semimajor axis shown by the horizontal line and focus (center of mass) shown by a plus sign. The primary (secondary) eclipses occur when the primary is located at the position marked by a square (diamond). The observer views these from a line of sight from the lower left (along the conjunction line from diamond to square). The longitude of periastron $\omega$ is measured from the ascending node crossing the plane of the sky (dotted line through center of mass position) to the periastron position at right. The true anomaly $\nu$ at primary eclipse is indicated as the angle from periastron to stellar position. Dotted lines show normals from the semimajor axis drawn through the eclipse positions out to the auxiliary circle inscribing the ellipse. The angle from periastron through ellipse center to the position on the auxiliary circle is the eccentric anomaly $E$, which is indicated for both eclipses. The solution for angle $2x$ is derived in the text.
\label{figa1}}
\end{figure}

\clearpage

\setcounter{figure}{0}
\renewcommand{\thefigure}{B\arabic{figure}}

\begin{figure} 
\begin{center} 
{\includegraphics[angle=0,height=12cm]{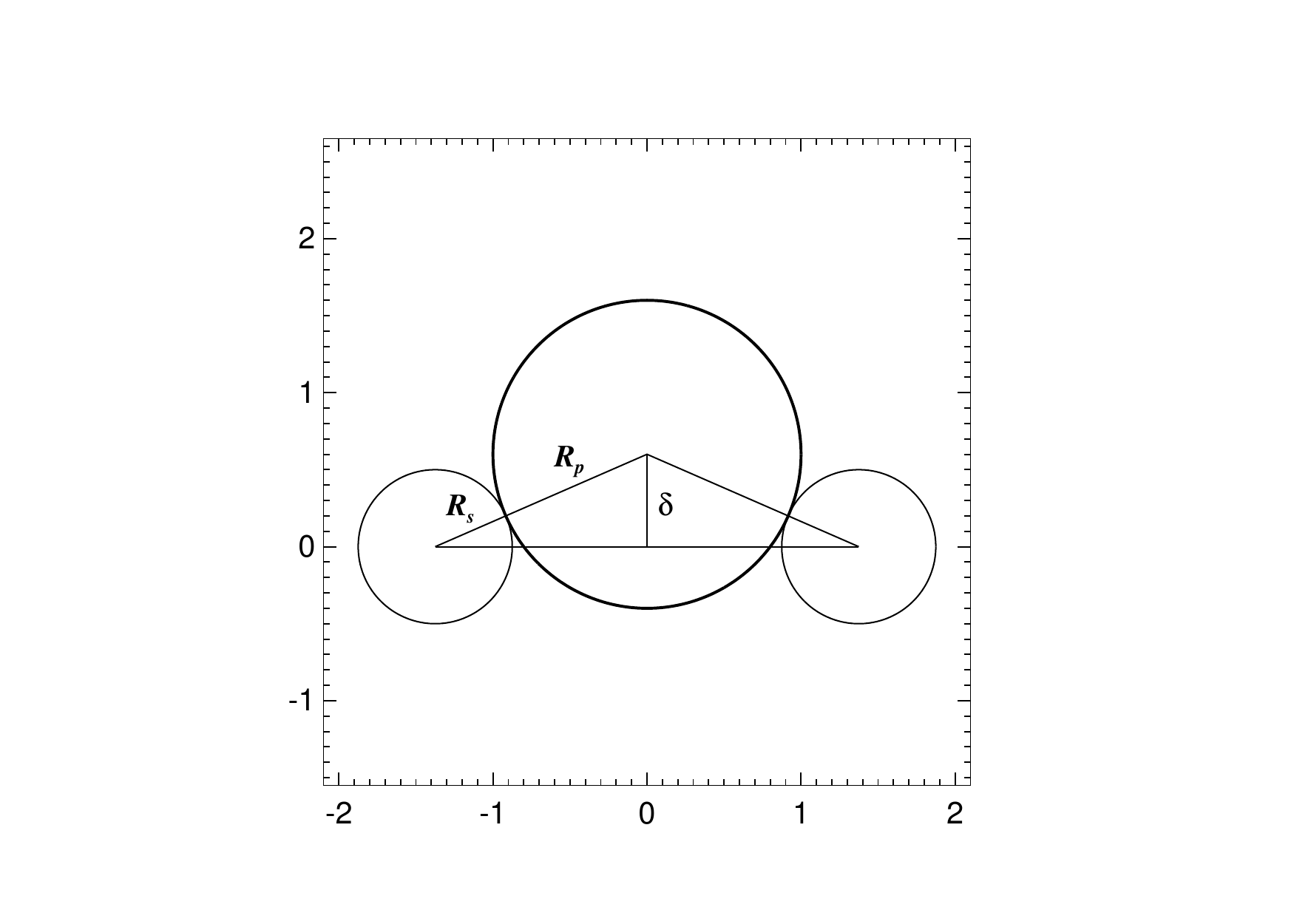}}
\end{center} 
\caption{A depiction of the appearance of an eclipse in the plane of the sky in the frame of the primary star.  The smaller secondary star moves from left to right attaining a minimum separation of projected centers indicated by $\delta = r \cos i$. The horizontal line connecting the center of the secondary at the start and end of the eclipse marks the projected distance of secondary motion. 
\label{figb1}} 
\end{figure} 
 
 \clearpage
 
\begin{figure} 
\begin{center} 
 {\includegraphics[angle=90,height=12cm]{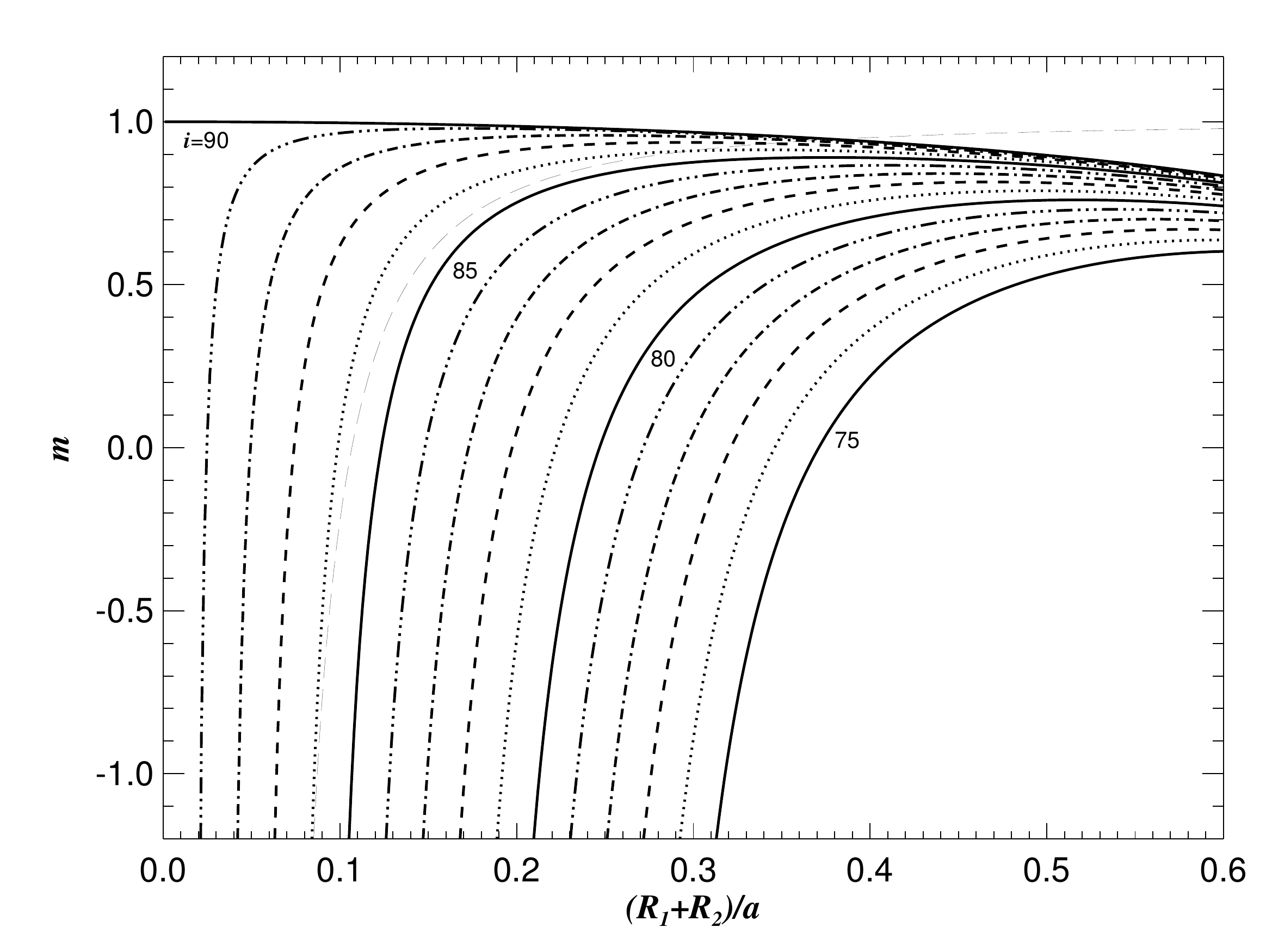}}
\end{center} 
\caption{The variation of the slope $m$ for the relation ${{d_s-d_p}\over{d_s+d_p}} = m ~ e \sin \omega$.  The values of the slope $m$ represent linear fits using equations B6 and B7 for small eccentricity for a range in assumed values of the relative sum of the radii $(R_1 + R_2 )/a$ and inclination $i$. Relations for constant $i$ are shown for $i=75^\circ$ to $i=90^\circ$ (bottom to top) in steps of $1^\circ$. The thin, long-dashed line in the middle represents the first order approximation from equation B4 for the case of $i=85^\circ$.   
\label{figb2}} 
\end{figure}


\begin{deluxetable}{cccccccc}
\tabletypesize{\scriptsize}
\tablewidth{0pt}
\tablenum{1}
\tablecaption{KIC 5738698 Radial Velocity Measurements\label{tbl-1}}
\tablehead{
\colhead{Date}                       & 
\colhead{Orbital}                    & 
\colhead{$V_{1}$}                  & 
\colhead{$\sigma_{1}$}          &
\colhead{$(O-C)_{1}$}           & 
\colhead{$V_{2}$}                  & 
\colhead{$\sigma_{2}$}          &
\colhead{$(O-C)_{2}$}            \\
\colhead{(HJD$-$2,400,000)} & 
\colhead{Phase\tablenotemark{a}}  &
\colhead{(km s$^{-1}$)}          &
\colhead{(km s$^{-1}$)}          &
\colhead{(km s$^{-1}$)}          &
\colhead{(km s$^{-1}$)}          &
\colhead{(km s$^{-1}$)}          &
\colhead{(km s$^{-1}$)}
}
\startdata
55366.7614 \dotfill &  0.297 &
\phn         $ -74.32$ &  1.18 & \phn\phs $   0.78$ &
\phn\phs     $  94.93$ &  1.79 & \phn\phs $   0.70$ \\
55368.7913 \dotfill &  0.719 &
\phn\phs     $  98.39$ &  1.23 & \phn\phs $   6.30$ &
\phn         $ -74.10$ &  1.68 & \phn\phs $   5.53$ \\
55368.8630 \dotfill &  0.734 &
\phn\phs     $  88.86$ &  1.30 & \phn     $  -4.42$ &
\phn         $ -86.03$ &  2.03 & \phn     $  -5.06$ \\
55368.9200 \dotfill &  0.746 &
\phn\phs     $  95.19$ &  1.35 & \phn\phs $   1.51$ &
\phn         $ -82.23$ &  1.76 & \phn     $  -0.75$ \\
55368.9649 \dotfill &  0.755 &
\phn\phs     $  89.37$ &  1.16 & \phn     $  -4.29$ &
\phn         $ -83.57$ &  1.77 & \phn     $  -2.04$ \\
55402.9316 \dotfill &  0.819 &
\phs \nodata  & \nodata & \phs \nodata &
\phn         $ -78.10$ &  3.91 & \phn     $  -4.29$ \\
55431.7764 \dotfill &  0.817 &
\phn\phs     $  87.15$ &  1.27 & \phn\phs $   1.00$ &
\phn         $ -72.22$ &  1.80 & \phn\phs $   1.96$ \\
55431.8614 \dotfill &  0.835 &
\phn\phs     $  82.60$ &  1.44 & \phn\phs $   0.85$ &
\phn         $ -68.73$ &  2.04 & \phn\phs $   0.99$ \\
55731.8001 \dotfill &  0.208 &
\phn         $ -73.96$ &  1.15 & \phn\phs $   1.89$ &
\phn\phs     $  96.83$ &  1.70 & \phn\phs $   2.48$ \\
55734.7638 \dotfill &  0.824 &
\phn\phs     $  84.21$ &  1.10 & \phn     $  -0.26$ &
\phn         $ -73.35$ &  1.50 & \phn     $  -0.87$ \\
55813.7096 \dotfill &  0.241 &
\phn         $ -80.02$ &  1.22 & \phn     $  -1.31$ &
\phn\phs     $  95.82$ &  1.62 & \phn     $  -1.75$ \\
55813.8211 \dotfill &  0.265 &
\phn         $ -79.34$ &  1.15 & \phn     $  -0.86$ &
\phn\phs     $  97.61$ &  1.50 & \phn\phs $   0.11$ \\
55813.8798 \dotfill &  0.277 &
\phn         $ -78.44$ &  1.28 & \phn     $  -0.82$ &
\phn\phs     $  95.34$ &  1.73 & \phn     $  -1.37$ \\
\enddata
\tablenotetext{a}{Relative to T$_0$ at primary eclipse.}
\label{tab:rvs}
\end{deluxetable}

\clearpage

\begin{deluxetable}{p{3.3cm}cc}
\tabletypesize{\scriptsize}
\tablewidth{0pt}
\tablenum{2}
\tablecaption{Orbital Solutions for KIC 5738698\label{tbl-2}}
\tablehead{
\colhead{Element} &
\colhead{Spectroscopic Solution} &
\colhead{ELC Solution}
}
\startdata
$P$~(days) \dotfill & 4.80877396\tablenotemark{a} & 4.80877396\tablenotemark{a} \\
$T_{0}$ (HJD$-$2,400,000)\tablenotemark{b}\dotfill & $55692.33\pm0.03$ &
     $55692.3348$\tablenotemark{a} \\
$e$ \dotfill & $0$\tablenotemark{a} & $0.0006$\tablenotemark{a} \\
$\omega$ (deg) \dotfill & \nodata & $52$\tablenotemark{a} \\ 
$K_{1}$ (km s$^{-1}$) \dotfill & $86.3\pm0.9$ & $86.2\pm$0.6 \\
$K_{2}$ (km s$^{-1}$) \dotfill & $89.7\pm0.9$ & $89.7\pm$0.9 \\
$\gamma$ (km s$^{-1}$) \dotfill & $ 7.8\pm0.6$ & $7.6\pm0.5$ \\
$M_{2}/M_{1}$ \dotfill & $0.96\pm0.01$ & $0.96\pm0.01$ \\ 
$a \sin i$ (R$_{\odot}$) \dotfill & $16.7\pm0.1$ & $16.7\pm0.1$ \\
\enddata
\tablenotetext{a}{Fixed.}
\tablenotetext{b}{Time of primary eclipse.}
\label{tab:sporb}
\end{deluxetable}

\begin{deluxetable}{p{3.2cm}cc}
\tabletypesize{\scriptsize}
\tablewidth{0pt}
\tablenum{3}
\tablecaption{Atmospheric Parameters from Reconstructed Spectra \label{tbl-3}}
\tablehead{
\colhead{Parameter} &
\colhead{Primary} &
\colhead{Secondary} 
}
\startdata
T$_{\mathrm{eff}}$ (K) \dotfill & $6792\pm50$ & $6773\pm50$ \\
$\log g$ (cgs) \dotfill & 4.05\tablenotemark{a} & 4.09\tablenotemark{a} \\
$V \sin i $ (km s$^{-1}$) \dotfill & $18\pm16$ & $21\pm10$ \\
$F_{2}/F_{1}$ ($\sim4275$\AA) \dotfill & \multicolumn{2}{c}{$0.82\pm0.06$} \\
$\log Z$ (cgs) \dotfill & \multicolumn{2}{c}{$-0.4\pm0.1$}
\enddata
\tablenotetext{a}{Fixed from light curve solution.}
\label{tab:atm}
\end{deluxetable}

\begin{deluxetable}{p{3.5cm}cc}
\tabletypesize{\scriptsize}
\tablewidth{0pt}
\tablenum{4}
\tablecaption{Combined Long Cadence Light Curve Fitting Parameters\label{tbl-4}}
\tablehead{
\colhead{Parameter} &
\colhead{Circular} &
\colhead{Eccentric} \\
\colhead{} &
\colhead{Solution} &
\colhead{Solution} 
}
\startdata
$T_{0}$ (HJD$-$2,400,000) \dotfill & $55692.335\pm0.003$ & $55692.3348\pm0.0004$ \\
$e$ \dotfill & 0\tablenotemark{a} & $0.0006\pm0.0003$ \\
$\omega$ (deg) \dotfill & \nodata & $52\pm23$ \\
$i$ (deg) \dotfill & $86.32\pm0.03$ & $86.33\pm0.03$ \\
$R_{1}$/a \dotfill & $0.1105\pm0.0001$ & $0.1097\pm0.0007$ \\
$R_{2}$/a \dotfill & $0.1023\pm0.0001$ & $0.1027\pm0.0008$ \\
$T_{2}/T_{1}$ \dotfill & $0.9922\pm0.0002$ & $0.9920\pm0.0003$ \\
\tableline
{\it{Kepler}} contamination \dotfill & 0.015\tablenotemark{a} & 0.015\tablenotemark{a} \\
Albedo (star 1) \dotfill & 0.5\tablenotemark{a} & 0.33 \\
Albedo (star 2) \dotfill & 0.5\tablenotemark{a} & 0.33 \\
$T_{\mathrm{grav}}$ (star 1) \dotfill & 0.068 & 0.068 \\
$T_{\mathrm{grav}}$ (star 2) \dotfill & 0.068 & 0.068 \\
\tableline
$T_{1}$ (K) \dotfill & $6792$\tablenotemark{a} & $6792$\tablenotemark{a} \\  
$T_{2}$ (K )\dotfill & $6739\pm51$ & $6740\pm52$ \\
$M_{1}$ ($M_{\odot}$) \dotfill & $1.39\pm0.04 $ & $1.39\pm0.04$ \\
$M_{2}$ ($M_{\odot}$) \dotfill & $1.34\pm0.06 $ & $1.34\pm0.06$ \\
$R_{1}$ ($R_{\odot}$) \dotfill & $1.85\pm0.02 $ & $1.84\pm0.03$ \\
$R_{2}$ ($R_{\odot}$) \dotfill & $1.71\pm0.02$ & $1.72\pm0.03$ \\
$a$ ($R_{\odot}$) \dotfill & $16.8\pm0.1 $ & $16.8\pm0.1$\\
$\log g_{1}$ (cgs) \dotfill & $4.0462\pm0.0004$ & $4.0525\pm0.0004$ \\
$\log g_{2}$ (cgs) \dotfill & $4.0963\pm0.0006$ & $4.0933\pm0.0006$
\enddata
\tablenotetext{a}{Fixed.}
\label{tab:elc}
\end{deluxetable}

\begin{deluxetable}{cccccccccc|cc}
\tabletypesize{\scriptsize}
\rotate
\tablewidth{0pt}
\tablenum{5}
\tablecaption{Long Cadence Segments \& Short Cadence Parameters\label{tbl-5}}
\tablehead{
\colhead{Parameter} &
\colhead{Q0-Q2} &
\colhead{Q3-Q4} &
\colhead{Q5-Q6} &
\colhead{Q7-Q8} &
\colhead{Q9-Q10} &
\colhead{Q11-Q12} &
\colhead{Q13-Q14} &
\colhead{Q15-Q17} &
\colhead{Short Cad.} &
\colhead{Mean} &
\colhead{Std Dev}  
}

\startdata
$T_{0}$ (HJD$-$2,455,692) \dotfill & 0.3347 & 0.3349 & 0.3348 & 0.3349 & 0.3336 & 0.3349 & 0.3348 & 0.3347 & 0.3348 & 0.3347 & 0.0004 \\
$e$ \dotfill & 0.0009 & 0.0006 & 0.0005 & 0.0004 & 0.0008 & 0.0010 & 0.0011 & 0.0009 & 0.0005 & 0.0007 & 0.0003 \\
$\omega$ (deg) \dotfill & 301 & 54 & 46 & 17 & 340 & 69 & 288 & 62 & 325 & 314/50 & 23/20 \\
$i$ (deg) \dotfill & 86.26 & 86.36 & 86.33 & 86.31 & 86.29 & 86.32 & 86.33 & 86.34 & 86.33 & 86.32 & 0.03 \\
$R_{1}$/a \dotfill & 0.1099 & 0.1116 & 0.1108 & 0.1095 & 0.1102 & 0.1100 & 0.1109 & 0.1106 & 0.1099 & 0.1104 & 0.0007 \\
$R_{2}$/a \dotfill & 0.1030 & 0.1006 & 0.1018 & 0.1035 & 0.1027 & 0.1027 & 0.1016 & 0.1019 & 0.1027 & 0.1023 & 0.0008 \\
$T_{2}/T_{1}$ \dotfill & 0.9920 & 0.9923 & 0.9925 & 0.9919 & 0.9927 & 0.9921 & 0.9921 & 0.9923 & 0.9920 & 0.9923 & 0.0003 \\
\enddata
\label{tab:segs}
\end{deluxetable}

\clearpage

\begin{deluxetable}{ccccccc|ccc}
\tabletypesize{\scriptsize}
\rotate
\tablewidth{0pt} 
\tablenum{6}
\tablecaption{Evolutionary Model Details \label{tbl-6}}
\tablehead{
\colhead{Model} &
\colhead{Enrichment Law} &
\colhead{} &
\colhead{Solar Composition} &
\colhead{} &
\colhead{Mixing Length} &
\colhead{He Diffusion} &
\colhead{Best-fit} &
\colhead{Best-fit} &
\colhead{Unscaled} \\
\colhead{} &
\colhead{(Y=)} &
\colhead{(X)} &
\colhead{(Y)} &
\colhead{(Z)} &
\colhead{($l/H_p$)} &
\colhead{} &
\colhead{$\log Z/Z_{UV_\odot}$} &
\colhead{Age (Gyr)} &
\colhead{Chi-Squared}
}
\startdata
Yonsei-Yale\tablenotemark{1} & $0.23+2Z$ & $0.7156$ & $0.2662$ & $0.0181$ & $1.7432$ & Y & $-0.31\pm0.08$ & $2.32\pm0.02$ & 0.5 \\

Victoria-Regina\tablenotemark{2} & $0.2354+2.2Z$ & $0.7044$ & $0.2768$ & $0.0188$ & $1.90$ & N & $-0.30\pm0.02$ & $2.16\pm0.01$ & 1.8 \\
 
PARSEC\tablenotemark{3} & $0.2485+1.78Z$ & $0.7343$ & $0.2756$ & $0.0152$ & $1.7$ & Y\tablenotemark{5} & $-0.23\pm0.04$ & $2.18\pm0.01$ & 3.6 \\

Geneva\tablenotemark{4} & $0.248+1.2857Z$ & $0.7200$ & $0.2660$ & $0.0140$ & $1.6467$ & Y\tablenotemark{6} & $-0.37\pm0.07$ & $2.3\pm0.1$ & 22.3 
  
\enddata
\tablenotetext{1}{Solar Mixture Source: \citealt{Grevesse:1996a}}
\tablenotetext{2}{Solar Mixture Source: \citealt{Grevesse:phys-scr1993a}}
\tablenotetext{3}{Solar Mixture Source: \citealt{Grevesse:aap1999a}}
\tablenotetext{4}{Solar Mixture Source: \citealt{Asplund:2005a}}
\tablenotetext{5}{For $M_{conv}>0.05 M_{tot}$}
\tablenotetext{6}{For $M<1.1M_{\odot}$}
\label{tab:evolmod}
\end{deluxetable}



\end{document}